\documentclass[conference]{IEEEtran}
\IEEEoverridecommandlockouts
\usepackage{amsmath,amssymb,amsfonts, verbatim, pdfpages}
\usepackage{graphicx}
\usepackage{hyperref}

\usepackage{biblatex}
\newcommand{\pvar}[1]{$p_{#1}\,$}
\newcommand{\pdef}{$p_{\text{defect}}\,$}
\newcommand{\pmu}{$p_{\mu}\,$}
\newcommand{\pmuout}{$p_{\mu-out}\,$}
\newcommand{\psigma}{$p_{\sigma}\,$}
\newcommand{\qvar}[1]{$q_{#1}\,$}

\newcommand{\ler}[1]{$\varepsilon_{\text{#1}}$ }

\def\BibTeX{{\rm B\kern-.05em{\sc i\kern-.025em b}\kern-.08em
    T\kern-.1667em\lower.7ex\hbox{E}\kern-.125emX}}

\begin{document}
\title{Boundaries of Acceptable Defectiveness: \\ Redefining Surface Code Robustness under Heterogeneous Noise}

\author{
    \IEEEauthorblockN{Jacob S. Palmer}
    \IEEEauthorblockA{\textit{Computer Science} \\
    \textit{Northwestern University}\\
    Evanston, U.S.A. \\
    JacobPalmer2026@u.northwestern.edu}
    \textit{Corresponding Author}
    \and
    \IEEEauthorblockN{Kaitlin N. Smith}
    \IEEEauthorblockA{\textit{Computer Science} \\
    \textit{Northwestern University}\\
    Evanston, U.S.A. \\
    kns@northwestern.edu}
}

\maketitle

\begin{abstract}
A variety of past research on superconducting qubits shows that these devices exhibit considerable variation and thus cannot be accurately depicted by a uniform noise model. To combat this often unrealistic picture of homogeneous noise in quantum processors during runtime, our work aims to define the boundaries of acceptable defectiveness (BADs), or the upper boundary of a qubit's physical error, past which this defective qubit entirely degrades the logical computation and should be considered faulty and removed from the surface code mapping. Here, we present a simulation framework based on the stabilizer simulation package Stim, that allows for rapid experimentation of quantum error correction (QEC) performance under any arbitrary and unique noise model. Using this tool, QEC circuits using rotated surface codes were generated, sampled, and analyzed from distances 3 to 17 with various defective error rates. In addition, we simulated heterogeneous noise models using the same parameters to test how increasingly deviated distributions of physical errors scale across code distances under realistic, heterogeneous noise models that are informed by current superconducting hardware. The results suggest that there are, in fact, boundaries of acceptable defectiveness in which a defective qubit, with a physical error rate $\leq 0.75$, can be left in the lattice with negligible impact on logical error rate given sufficient code distances and proper placement in the lattice. Additionally, when modeling noise as a uniform distribution, the logical error rate shows minimal impact with increasing deviation, with $\sigma \, \leq \, \mu$. As a result, we propose that defectiveness of both individual qubits and the overall uniformity of lattice fidelity should not be viewed as all or nothing, but instead as a spectrum. Our research demonstrates how heterogeneity relates to the logical error rate and, through the framework provided, facilitates the development of preliminary goals and metrics for hardware designers to achieve a target logical performance with imperfect or non-uniform qubit qualities.
\end{abstract}

\section{Introduction} {
    Quantum error correction (QEC) enables fault-tolerant quantum computing by utilizing multiple physical qubits to redundantly encode a single abstract logical qubit, thereby making the information more resilient to errors such as decoherence or gate noise \cite{fowler_surface_2012}. A leading method for QEC is the surface code (SC), in which physical qubits are arranged in a 2D lattice, and operations are performed between neighboring qubits~\cite{fowler_surface_2012}. The SC is favorable for near-term QEC demonstrations because of its high noise tolerance, and defective qubits in the 2D lattice can be managed by deforming the lattice around the defect at the cost of additional physical qubit overhead~\cite{auger_fault-tolerance_2017, strikis2023quantum}. Existing SC research explores (1) how logical error rate (LER) is affected by physical error rate and (2) how physical overheads are impacted by defects. However, analysis often considers physical qubit performance, specifically in regard to superconducting qubit architectures, in a rigid, homogeneous manner wherein physical qubits are either characterized as either uniform in their error rates or completely unusable. As a result, critical insights into the impact and upper bounds of the usability of qubits that range in their defectiveness are currently sparse. A better understanding of how SC tolerates heterogeneous error landscapes is necessary to improve the mapping of QEC to realistic quantum systems.
      
    The primary goal of this work is to better understand how the SC responds when the physical error rate is not held constant across a lattice, such as with normally distributed noise or individual defects, and to provide a framework for analyzing performance in the presence of heterogeneity. Using a QEC simulation framework that supports distributed error models and is based on Stim, we investigate the impacts of defective outlier qubits and normally distributed noise on overall LER across code distances ranging from 3 to 17. Our contributions are as follows:
    
        \begin{itemize}
            \item For the first time, we define the concept of \textit{boundaries of acceptable defectiveness }(BADs) as a way to examine and quantify the performance and thresholds for imperfect noise models. The community lacks quantitative guidance with respect to the question of ``how defective is too defective'' for qubit error distance from the mean. Our research shows the utility of treating the physical error rate as a distribution when selecting the best SC solution to achieve an algorithm's target logical error. Further, our simulation results show (1) the existence of BADs in both homogeneous and heterogeneous noise models and (2) these BADs are not static and differ greatly depending on the assumed noise model (i.e., the quality of qubits within a lattice) and targets for LER. Real quantum hardware exhibits vast variation in noise, and thus, in the pursuit of grounding our simulated models in reality, identifying the BADs allows a more accurate understanding of how performance scales given any arbitrary noise model. 
            \item We developed a novel circuit generator framework using Stim for rapid experimentation of SC performance under any arbitrary and unique noise model. Our code, which will be open source, creates quantum circuits with granular control over each qubit's fidelity, enabling comparisons across any configuration of noise models. It also provides a novel generation of heterogeneous noise models based on deviations from a mean physical error rate. This tool has a high potential to find utility in QEC resource estimation and design space exploration based on superconducting qubits. This code and demo notebooks used to generate the figures within this paper can be found \href{https://github.com/JacobSPalmer/quantum-ddq-toolkit}{here on GitHub}.
            \item We model the performance of the SC under heterogeneous noise models, providing additional insights into the relationship between logical performance on imperfect quantum devices and increasingly deviant physical noise distributions. 
        \end{itemize}
}
\section{Background} {
\label{sec-background}
    \subsection{The Surface Code}{
        \begin{figure}[t!]
            \centering
            \includegraphics[width=0.9\linewidth,trim={0cm 0cm 0cm 0cm},clip]{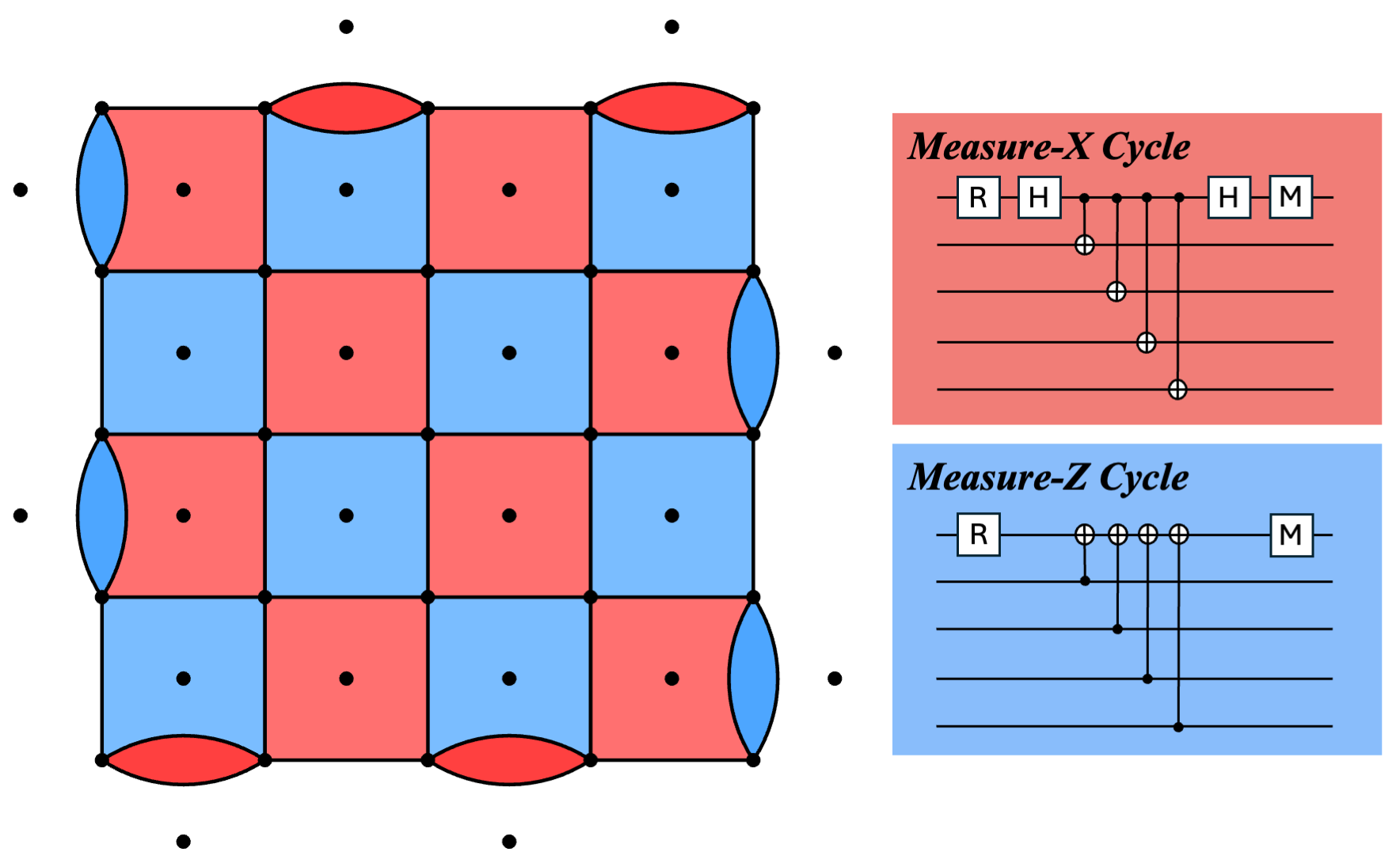}
            \caption{An example visualization of a d=5 rotated surface code lattice (left) where data qubits are found on the grid boundaries, and X (Z) measurement/stabilizer qubits are found within the red (blue) squares of the grid. The sequence of operations (right) for a single Measure-X and Measure-Z surface code cycle are also illustrated.}
            \label{fig-d5-sc-and-circuits}
        \end{figure}
    
        The surface code~\cite{fowler_surface_2012} (SC) is one of the most promising QEC codes for physical realization, especially with superconducting qubits, due to its implementation using a 2D nearest-neighbor qubit layout, and its high tolerance to noise~\cite{fowler_surface_2012}. Recent demonstrations on superconducting circuits have allowed logical qubits to be demonstrated and LERs to drop below threshold~\cite{acharya_quantum_2025}. SCs can implement a universal set of gates, create logical entanglement with lattice surgery~\cite{litinski2019game}, and generate magic resource states with magic state distillation~\cite{litinski2019magic,gidneyEfficientMagicState2019}. In this work, we focus on the rotated SC (i.e., any further mention of SC is assumed to be this), requiring $2d^2-1$ physical qubits per logical qubit. Our goal is to study the SC's ability to store logical qubit states over time while being subject to various noise models.
    
        An illustration of a d=5 SC patch can be seen on the left side of Fig.~\ref{fig-d5-sc-and-circuits}. In this picture, $d^2$ data qubits are found on the edges of the grid while $d^2-1$ measurement qubits are found within the red and blue squares representing Measure-X and Measure-Z stabilizers, respectively. The right side of Fig.~\ref{fig-d5-sc-and-circuits} shows the sequence of operations performed during the SC cycle that allows the data qubits (bottom four lines) to interact with the measurement qubit (top line). Over many cycles, the history of detection events gives insight into the errors happening during runtime. These syndromes are then used by the decoding algorithm, minimum-weight perfect matching (MWPM)~\cite{dennis2002topological} in our case, to locate the error within the logical qubit so that it can be corrected. A logical error is introduced into the SC when at least $d/2$ errors occur along a non-trivial path.
    }
    \subsection{Previous Work} {
        The heterogeneous nature of superconducting qubit physical qubit error has been well-documented~\cite{carrollSubsystemSurfaceCompass2024, murali2019noise,dasgupta2023reliable}. Prior work has proposed novel architectures, such as those based on chiplets~\cite{smith2022scaling}, to promote device scaling with more uniform metrics, but it is unlikely that quantum systems comprised of imperfectly fabricated processors will ever demonstrate perfectly homogeneous noise. This raises many challenges in terms of how to design QEC strategies that best meet the needs of algorithms while managing realistic constraints of hardware.
        
        Previous work with the SC has made progress in combating the reality of defective qubits by leveraging “super stabilizers” and boundary deformations in the SC patch. These techniques work by routing the stabilizer measurement circuits so that they avoid faulty components~\cite{lin_codesign_2024,debroy_luci_2024,yin_surf-deformer_2024}. Current deformation methodologies incur a tradeoff: deform the lattice around the defect, thus removing the defect's negative impact on the LER \textit{but} incur additional resource overhead in the process. This increase in overhead is typically due to a higher physical qubit requirement, but some deformation techniques also incur increased complexity during compilation or runtime through specialized steps to facilitate computation within the deformed lattice. 
        
        Most work in SC defect management assumes a rigid categorization of qubit quality in the sense that a qubit is either operating at the mean physical error rate or it is completely unusable and should be dropped out of the logical qubit mapping. However, more lenient bounds of acceptable defectiveness could exist, especially in the case where the distribution of physical qubit error rates is characterized by a standard deviation. For example, the work in~\cite{mohseni_how_2025} reports that the current variation in coherence time seen on existing superconducting qubits leads to hindered LER with the SC. This presents the question of how defects should be defined, and which qubits with higher error rates can be left in the lattice in its current state, given the QEC designer's choice of code distances. These sorts of estimates remain to be fully explored, with early efforts appearing in Carroll et al. \cite{carrollSubsystemSurfaceCompass2024}, which examined how logical error rates for subsystem surface and compass codes on a heavy-hex lattice depend on non-identical infidelity distributions and the presence of "bad" outlier qubits. Our work validates and extends this type of analysis for rotated surface codes, and uses it as a foundation for developing the boundaries of acceptable defectiveness in the presence of non-uniform qubit fidelity.

        \begin{comment}
        as such, this work serves as a starting point for scaffolding out potential defects, their impacts, and the design choices necessary to minimize detrimental effects. To narrow our scope in this initial exploration of defining boundaries of acceptable defectiveness, we focus on distributions associated with physical qubit error (Section~\ref{sec-background}). 
        \end{comment}
    }
}
\section{Homogeneous vs. Heterogeneous Assumptions of Noise on Surface Code Lattices} {
\label{sec-homogeneous-vs-heterogeneous-assumptions}
    In general, if physical operations on the SC are performed below a sufficiently low critical noise threshold then logical error rates are exponentially suppressed with increasing code distances \cite{fowler_surface_2012}. This scaling can be expressed by the relation:
    
    \begin{equation}
    \varepsilon \propto \left( \frac{p}{p_{thr}} \right)^{\frac{d+1}{2}}
    \end{equation}
    
    \noindent where $\varepsilon$ is LER, \pvar{thr} is threshold for physical noise tolerance, and \textit{d} is the distance of our code \cite{acharya_quantum_2025}. Prior work has found \pvar{thr} = 0.0057~\cite{fowler_surface_2012}. In this general form relation, \textit{p} represents the physical error rate of all $2d^2-1$ physical qubits that encode our singular logical qubit in a given lattice. Thus, this relation assumes uniform and homogeneous physical noise exhibited across the lattice. However, results from real quantum devices have shown that superconducting quantum devices do not exhibit uniform noise throughout the processor, but instead vary greatly in qubit fidelity \cite{mohseni_how_2025, tannu2019not}.
}
\section{Simulation Framework Overview} {
\label{sec-framework-overview}
    The tool developed as part of this research is, at its core, a quantum circuit generator with granular per-qubit physical noise control on rotated SCs using Stim \cite{gidney_stim_2021}. The circuits generated are memory experiments, in which the goal is to identify how well a logical observable is preserved over time. In essence, a memory experiment involves initializing a logical qubit in a known state, performing repeated rounds of syndrome extraction, and measuring the logical observable \cite{gidneyStabilityExperimentsOverlooked2022}. If no logical errors occur, the final measured logical observable at the end of the circuit will be the same as the initial logical observable. Memory experiments are a reliable benchmark for stabilizer codes, like the surface codes, in determining how logical computation is affected by characteristics like code distance and how physical noise affects our logical computation. Stim is specifically designed for rapid design and simulation of quantum stabilizer circuits, and the tool presented here uses Stim to generate dynamic noise memory experiments \cite{gidney_2025_15354872}. While other popular quantum circuit simulators exist \cite{javadi-abhariQuantumComputingQiskit2024, cirqdevelopersCirq2025} Stim was selected because (1) the ability to adaptively create entire stabilizer circuits from string manipulation and (2) it offers very fast runtime allowing for sampling the logical error count from millions of shots in minutes. For example, in our distributed heterogeneous noise simulations in Section \ref{subsec-case-3}, the simulations (Fig. \ref{fig-case3-multiplot-performance-0} and Fig. \ref{fig-case3-multiplot-performance-1}) consist of nearly $4*10^7$ total shots over the eight distances sampled for each, with each point representing a generated circuit with a unique noise distribution with $5*10^4$ shots. Thus, it was important that the simulation tool be reasonably fast in order to test a wide variety of different input parameters.
    
    The circuits are generated with a variety of input parameters to facilitate testing variations in both individual qubit noise (e.g., specific defect location and defect physical error rates) and overall lattice noise distributions (e.g., uniform or heterogeneous). Our code works in three parts (1) generating the parametrized Stim string representation of the Z-memory experiment circuit, (2) parsing the simulation data and performance results provided by Stim into usable DataFrames, and (3) directly plotting simulation data into easy-to-understand plots that allow for immediate comparison of performance across any number of code distances. Each of these circuits is built using either a uniform (homogeneous) or normally distributed / defect-injected (heterogeneous) noise distribution. 

    \subsection{Building Circuits as Stim Strings} {
        For a given distance \textit{d}, the generators first assign each of the $n=2d^2-1$ qubits a coordinate pair (X, Y) within a grid representing the 2D lattice. The set of values for both the x-coordinate and the y-coordinate starts at 0.5, increments by 0.5, and goes up to $(d+0.5)$. 
        
        In SCs, each qubit is either a measurement (stabilizer) qubit or a data qubit \cite{fowler_surface_2012}. Within our generators, data qubits are always located at integer coordinates (e.g., (3, 5)) and measurement qubits are always located at non-integer coordinates $\pm.5$ in x or y direction from the data qubits (e.g., (.5, .5) or (1.5, 2.5)). Within our setup, data qubit coordinates are always integer values, while the measure qubits will always be $\pm.5$ from their respective data qubits.

        \begin{figure}[!htb]
                \centerline{\includegraphics[width=0.8\columnwidth]{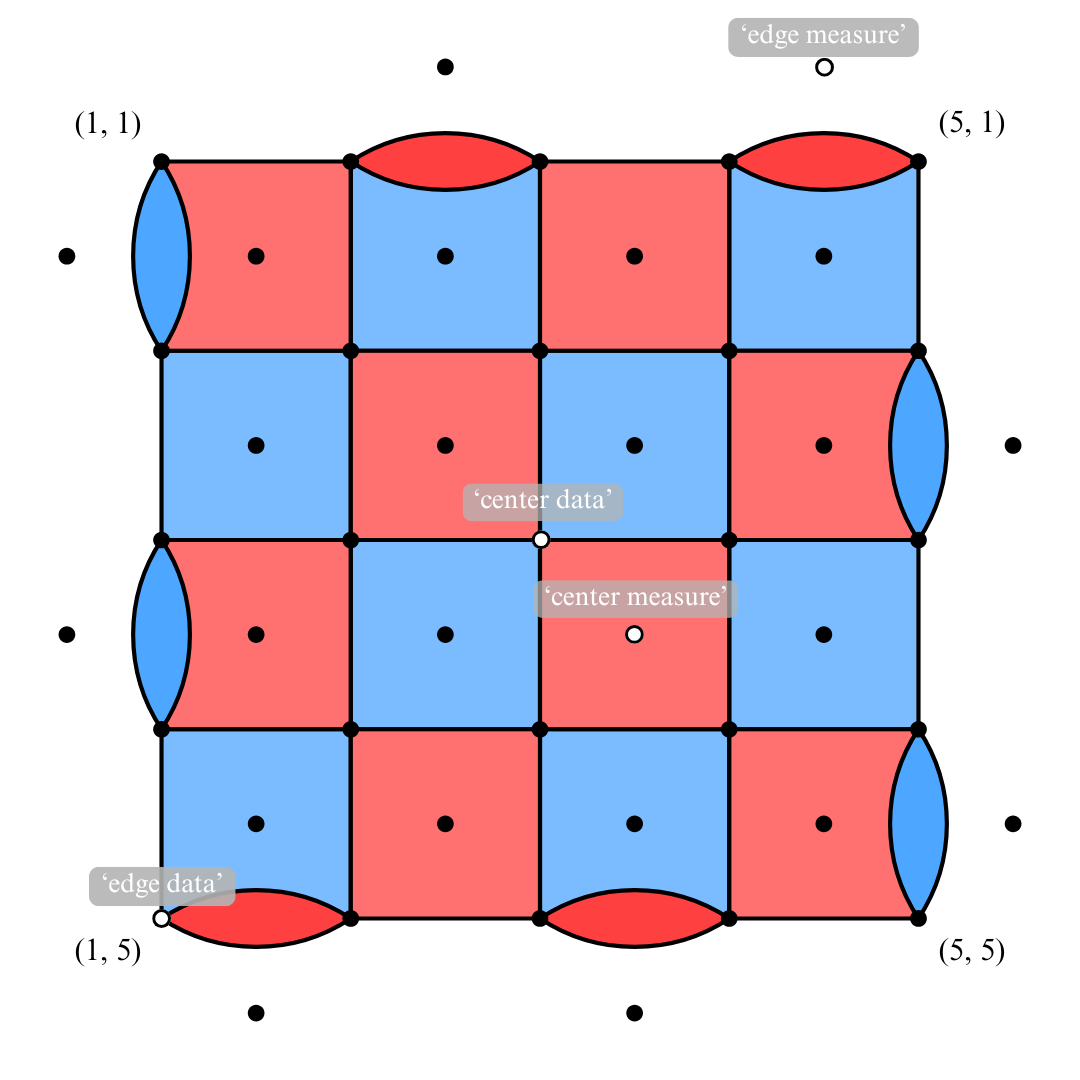}}
            \caption{ For the simulations and selection of defect locations, the generators optionally allow for specific qubit locations to be designated in order to maintain similarity across various code distances. The qubit locations are marked by the special case strings, 'center data’, 'center measure’, 'edge data', 'edge measure'. These special case strings are then converted by calculating their respective coordinate given any lattice of distance \textit{d}.
            }
            \label{fig-defect-keyword-grid}
        \end{figure}
        
        Because Stim strings represent the stabilizer circuits in a manner similar to quantum circuit diagrams \cite{nielsenQuantumComputationQuantum2010}, each of the \textit{n} qubits is referred to by a wire index from 0 to $n-1$. Our generators map coordinate pairs to the appropriate wire index within the Stim string, allowing for the generators to accept coordinates as inputs for specific noise models and then appropriately translate between the index and coordinate references for any given qubit. Additionally, the generators will map special location keywords to coordinates on the lattice, as seen in Fig. \ref{fig-defect-keyword-grid}. Since the distance directly determines the grid space, these special keywords facilitate placing defects at the same relative location in the lattice, simplifying cross-distance performance analysis. 
    }
    \subsection{Homogeneous Distributions}{
        The homogeneous generator takes as input a physical error rate \textit{p} that will be set as the standard physical error rate for all qubits in the lattice. This represents the idealized uniform assumption of noise across a lattice.
    }
    \subsection{Heterogeneous Distributions}{
        \label{subsec-heterogeneous-distributions}
        As discussed earlier, a primary motivation for this research is to provide methods for exploring, simulating, and analyzing noise models that more accurately resemble real superconducting devices \cite{mohseni_how_2025, acharya_quantum_2025}. To achieve this, the heterogeneous generator creates a circuit that can be used to model the performance of a non-perfect lattice with individual qubits' physical error rate spanning any arbitrary distribution of noise values, rather than a static \textit{p}. 
        
        In this first iteration of our code, the heterogeneous generator defaults to a truncated Gaussian distribution. For each qubit in a lattice, its physical error rate is randomly sampled from a truncated normal distribution created based on two input parameters: a mean physical error rate, \pmu, and either standard deviation, \psigma, or a scalar, $\alpha$, which determines \psigma as such:
        \begin{equation}
        p_\sigma = \alpha *p_{\mu}
        \end{equation}
        
        \begin{figure}[ht!]
            \centerline{\includegraphics[width=1\columnwidth]{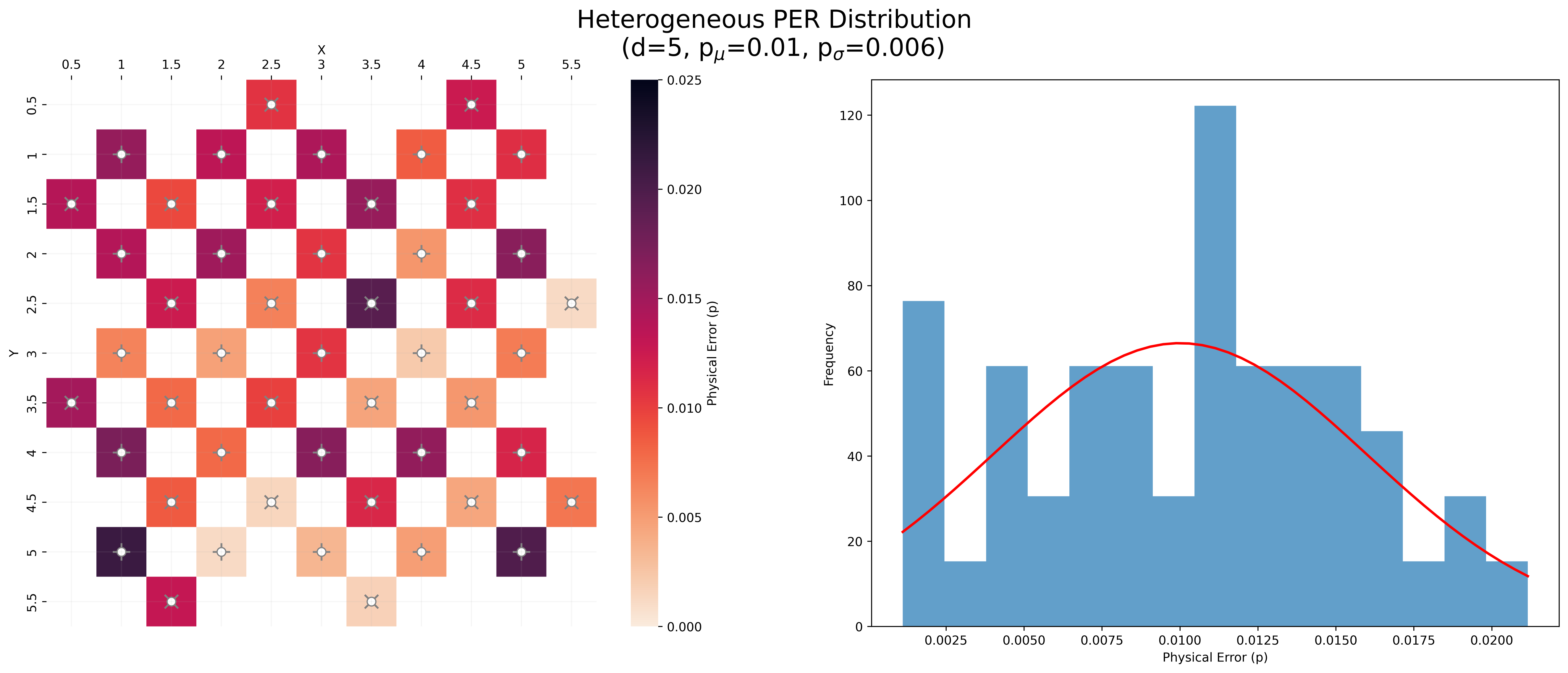}}
            \caption{ A heatmap and distribution representing a randomly distributed heterogeneous noise distribution on a d=5 code. The distribution on the right is shown to demonstrate that, although the truncated normal distribution results in slight differences, the samples still form a distribution around \pmu and $p_{\sigma}$. For our experiments in Case 3 (Section \ref{subsec-case-3}) and Case 4 (Section \ref{subsec-case-4}), we utilize an $\alpha$ parameter and the actual sampled distribution mean to ensure that any shifts from the input mean are reflected (i.e., when $p_{\sigma} > p_{\mu}$ at sufficiently small \pmu values). In this figure, with a $p_{\sigma}=0.006$ and $p_\mu=0.01$, the $\alpha$ parameter would be $\alpha=0.6$, given that $p_\sigma = \alpha*p_\mu$. Within all heatmaps shown, data qubits are denoted by \textbf{+} and measurement qubits are denoted with \textbf{X}.}
            \label{fig-heterogeneous-heatmap-with-pdf}
        \end{figure}
        
        Each time a circuit is generated with a given input \pmu and \psigma, a distribution following these parameters will be randomly sampled from, creating unique distributions of heterogeneous noise between circuit generations, as seen in Fig. \ref{fig-heterogeneous-heatmap-with-pdf} for a d=5 code. This unique distribution for each generated circuit is, in part, a modest reflection of how the noise exhibited by superconducting qubits can greatly fluctuate during or between runtime, but introduces some variability in our simulation's performance results that leads to some trends being non-monotonic, something that is discussed in Section \ref{sec-case-studies} and Section \ref{sec-future-work} later on.
        
        It is important to note that noise distribution on current superconducting devices does not necessarily follow a Gaussian model 
        \cite{sungNonGaussianNoiseSpectroscopy2019}. Thus, as a preliminary method for analyzing the impact of increasingly distributed noise in a lattice, analyzing how the SC reacts under a Gaussian distribution is an ideal starting point, and our parameterized modeling allows for other distributions to be explored. 
    }
    
    \subsection{Defective Qubits} {
        Both homogeneous and heterogeneous circuit generators accept a number of parameters, including distance, rounds, maximum shots. More importantly, the generator methods allow individual defect(s) to be arbitrarily created at any part in the lattice by providing a qubits coordinate (x, y) pair within the lattice (as seen in Fig. \ref{fig-defect-keyword-grid}) and a physical error rate for each defect, \pdef.
    }
    \subsection{Operational Noise}{
    \label{subsec-operational-noise}
        For this initial foray into heterogeneous noise models, we adopt a simplified notion of noise. Specifically, our memory experiments primarily use depolarizing-channel (D) noise, a standard quantum noise model in which a qubit is depolarized with probability $p$ \cite{tomitaLowdistanceSurfaceCodes2014}. In most contexts, $p$ is treated as the overall physical noise level of a qubit, encompassing more than just depolarization \cite{fowler_surface_2012}. In our simplified model, however, noise enters only through Pauli‑X (bit‑flip) errors on state preparation and measurement and depolarizing errors on $CX$ gate operations. In reality, quantum noise can be introduced by a wide variety of sources, such as interactions with the environment, uncontrolled interactions with other qubits, imperfect operations, or leakage~\cite{resch_benchmarking_2021}. Additionally, noise can be introduced permanently on specific qubits as a result of manufacturing defects on the superconducting devices \cite{auger_fault-tolerance_2017}. Implementing a more complex noise space is an area that could be greatly extended upon and is discussed further in our future work section.
        
        In our generators, noise is introduced right before circuit elements (e.g., gates and measurements) as a noise operator with a given probability of being applied, $p$. Each qubit is assigned an individual error rate, \pvar{i}, where \textit{i} is the index of the qubit. \pvar{i} determines the error rate for all operation involving this qubit. These operations include all multi-qubit operations (“DEPOLARIZE2”), single-qubit operations (“DEPOLARIZE1”), and measurements (“X-ERROR”) on the qubit. The \pvar{i} is scaled based on single or two-qubit operations. Generally, multi-qubit operations are more prone to errors than operations involving only a single qubit. Thus, it should be reflected that CX operations involving the defective qubits would exhibit a higher error rate than their single-qubit operations. Thus, for two-qubit depolarization noise associated with CX operations, the error rate is set as a mean of the \pvar{i} corresponding to the two qubits involved in the operation. Since two-qubit operations are more prone to error than single-qubit operations \cite{acharya_quantum_2025,mohseni_how_2025}, two-qubit noise has an additional scaling factor, \textit{s}, on the \qvar{i} + \qvar{i} mean, in the form:
        
        \begin{equation}
        p_{CX}=\left( \frac{p_{q1}+p_{q2}}{2} \right)*s 
        \end{equation}
        
        \noindent This scaling factor is set as \textit{s}=1.2, representing a 20\% increase over single qubit operations. For single-qubit depolarization and Pauli-X, the value for each operation on a qubit \qvar{i} is set directly to its corresponding \pvar{i} value. 
    }
}
\section{Case Studies and Analysis}{
    \label{sec-case-studies}
    To demonstrate the performance disparity between heterogeneous and homogeneous noise assumptions, we examined how LER scales across code distances 3-17 when assuming a homogeneous vs. heterogeneous noise model. In each case, the logical error rate per round, \ler{target}, is the metric used to evaluate performance. Every circuit generated for these memory experiments contains 3 rounds of syndrome extraction followed by a measurement in the Z-basis. A shot refers to one complete run of the circuit. The logical error rate for each shot is considered as the average number of errors that occur per round in the measurement circuit: 
    
    \begin{equation}
    \varepsilon_{round}=\left( \frac{errors}{shots*rounds} \right)
    \end{equation}
    
    \noindent To facilitate comparison and benchmarking across the four cases, we establish a target logical error rate, \ler{target}, of 0.005. The \ler{target} to use as a baseline depends primarily on the fidelity required to support a specific algorithm or application on a particular quantum device. With Shor's algorithm, for instance, past literature has estimated a \ler{target}$\in[10^{-15},10^{-14}]$ in order to reach sufficient fidelity to achieve an acceptable chance of success \cite{fowler_surface_2012}. For our purposes, we assume that a \ler{target} $=0.005$ is considered to be sufficient fidelity for the generated memory experiments. While a \ler{target} $=0.005$ is significantly higher than in practical use cases such as Shor's algorithm, it reflects a realistic logical error rate achievable with emerging hardware as we approach the early fault-tolerant era \cite{acharya_quantum_2025}. A primary benefit of our proposed BADs framework is that it enables performance analysis to contextualize targeted applications within given hardware configurations and constraints. As such, these case studies serve as both an analysis reflecting current estimates of superconducting hardware \textit{and} an example of how the BADs framework can be applied to benchmark performance for any given application. 
    
    In these experiments, we once again return to the idea of the \textit{“boundary of acceptable defectiveness”}, or BADs. For each code distance, the boundary of acceptable defectiveness (BAD) defines the maximum physical noise tolerance after which the LER surpasses the \ler{target}. Within the figures for each of our case experiments, the BAD for each code distance is precisely at the intersection point between the target \ler{target} and the distance's performance curve. Once the physical noise level has surpassed the BAD, we can consider the lattice to be effectively too noisy and the computation unusable. Broadly speaking, the BAD can be viewed as a function of a processors distance, unique noise characteristics, and the selected \ler{target}:
    
    \begin{equation*}
    BAD(p) = f(\text{distance}, \text{noise configuration}, \varepsilon_{\text{target}})
    \end{equation*}
    
    Importantly, these boundaries of acceptable defectiveness for a given distance are not held constant across different models of noise. Thus, using these boundaries provides a meaningful way to quantitatively compare performance across various noise models. Using our tool, the boundaries of acceptable defectiveness (BADs) are provided by determining the intersection point of the logical performance curve and the targeted logical error rate. More formally, we can define $I$ as the piecewise linear interpolant through the set of points for each sample, and let $L$ be the horizontal line representing the targeted logical error rate. Then, for a given sample, we define the $BAD$ as the solution of $L(x)=I(x)$.
    
    Four primary cases were selected under increasingly complex models of noise: (1) a completely uniform homogeneous lattice, (2) a uniform homogeneous lattice with a single defect in the center data qubit, (3) a heterogeneous lattice, and (4) a heterogeneous lattice with a single defect in the center data qubit. In Cases 2, 3, and 4, we vary a single case-specific parameter and measure performance across code distances, allowing the impact of that parameter on performance scaling with distance to become evident. For Cases 2 and 4, this parameter is the defective qubits physical noise \pdef; in Case 3, it is a deviation scalar $\alpha$ that determines how far the physical noise of all qubits deviates from the mean. For each distance and each value of the case-specific parameter, we generate 20 distinct physical error rates $p$ (mean physical error rates, \pmu, in Cases 3 and 4) logarithmically spaced between $10^{-3}$ and $10^{-1}$. For each combination of $d$, $p$ (\pmu), and case parameter, one circuit is constructed and the performance is simulated over 50,000 shots. The logical error rate plotted at each point is the average logical error rate per round over those 50,000 shots, so each curve shows how a single sampled processor with a fixed parameter behaves as the overall noise level varies.
    
    \subsection{Case 1: Homogeneous Noise} {
        \label{subsec-case-1}
        Case 1 covers the assumption of uniform noise in which all physical qubits within a lattice unilaterally express the same \textit{p} noise. A heatmap for this uniform homogeneous noise model can be seen in Fig. \ref{fig-case1-heatmap}. The performance results for Case 1 are shown in Fig. \ref{fig-case1-performance} for d=[3, 5, 7, 9, 11, 13, 15, 17]. 
            \begin{figure}[!ht]
            \centerline{\includegraphics[width=0.75\columnwidth]{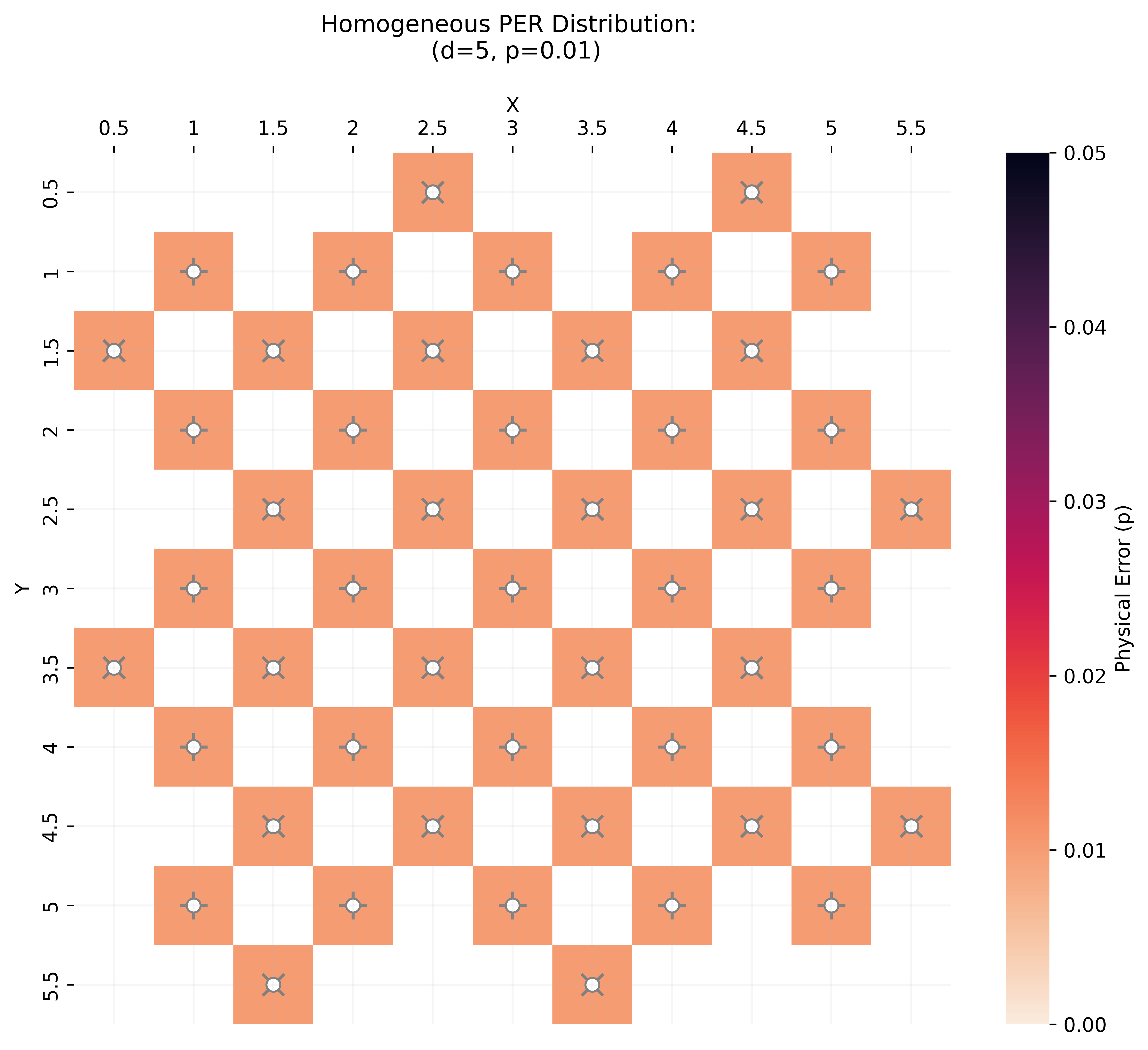}}
            \caption{ \textit{$($Case Study 1$)$} Heatmap for a homogeneous noise model on a d=5 code with no defects or deviations. In this heatmap, data qubits are denoted by \textbf{+} and measurement qubits are denoted with \textbf{X}.}
            \label{fig-case1-heatmap}
            \end{figure}
            \begin{figure}[!ht]
            \centerline{\includegraphics[width=0.8\columnwidth]{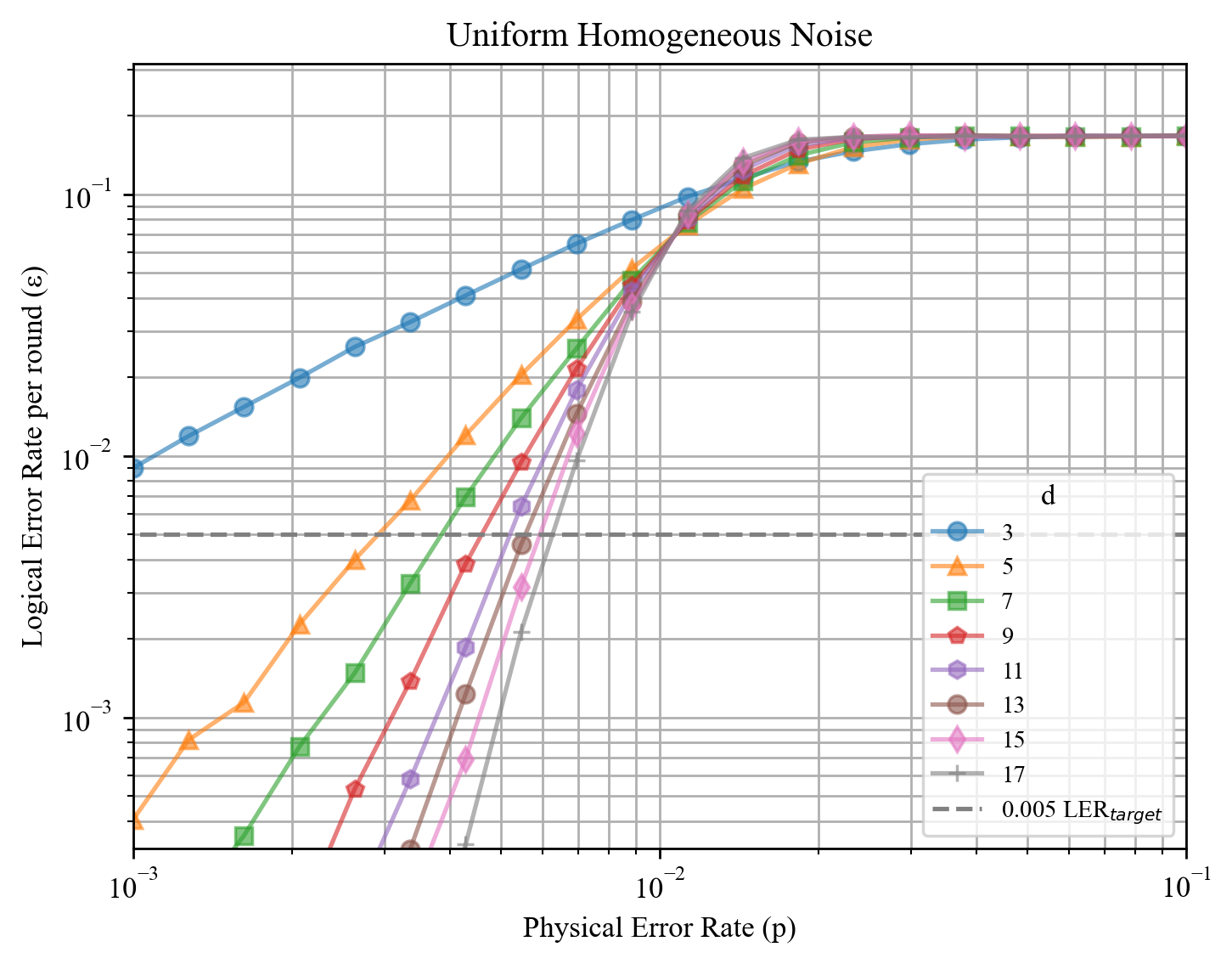}}
            \caption{ \textit{$($Case Study 1$)$} Performance results for a completely uniform, homogeneous noise model across distances d=[3, 5, 7, 9, 11, 13, 15, 17]. Each point in the plot represents a single simulated processor instance: for each fixed distance, we generate one circuit per $p\in\{10^{-3+(\frac{2*k}{19}})|k\in\{0, 1, 2, ..., 19\}\}$ and estimate the logical error rate at that point from 50,000 shots of that circuit. Case 1 represents the most ideal noise model assumptions when considering SCs in that all qubits within the lattice have an identical quality of \pvar{} and no qubit deviates from this value. In Cases 2 and 3, the blue line in each performance result figure is the baseline for each distance when no deviation is expected. As such, each line in this figure is functionally equivalent to the blue baseline in the Case 2 and Case 3 performance results for the corresponding distance. Because errors are probabilistically applied, small differences in these baselines can occur across simulations.}
            \label{fig-case1-performance}
            \end{figure}
        In past literature, this most closely resembles the ideal and common notion of how noise is represented in SC. As shown in Fig. \ref{fig-case1-performance}, all distances (with the exception of d=3) cross the \ler{target}. For the highest distance sampled, d=17, the BAD is at $p\approx0.006$. Any d=17 lattice with a $p\leq0.006$ would, in this theoretical model where \ler{target}$=0.005$, be sufficient for computation. However, if our goal is to minimize the code distance and thus our physical qubit overhead, we would select the minimum viable code distance for which our device's expressed \pvar{} is below the BAD for that distance. In this uniform homogeneous model, the d=5 code has a BAD at approximately $p\approx 0.003$. Thus, if deciding an optimal code distance for a physical device capable of performing computation with physical noise, $p\approx 0.003$, we should select the d=5 code. While achieving $p=0.003$ on a real superconducting device might be out of reach with current hardware \cite{kimErrorMitigationStabilized2025}, analyzing the BADs for a given code distance with a realistic physical noise model allows for additional optimization considerations. In Case 1, performance requirements are met at all but the smallest distance, d=3. While achieving uniform noise across all physical qubits results in the highest overall fidelity and the most viable code distances, noise data from superconducting devices exhibit significant variations in noise within a given chip. Thus, this uniform model in Case 1, at least for superconducting devices, is naive and an unrealistic basis for accurately modeling SC robustness \cite{kimErrorMitigationStabilized2025,mohseni_how_2025}.
    }
    \subsection{Case 2: Homogeneous Noise with Outlier Defect} {
    \label{subsec-case-2}
        To test individual qubit impact within a homogeneous noise model, two separate circuits were generated with a defect located at the center data qubit and near-center measure qubits. In this second scenario, we examine the impact of a single defect on the SC’s performance under an otherwise uniform noise model. A heatmap for the homogeneous noise model with an outlier defect center data qubit can be seen in Fig. \ref{fig-case2-heatmap}. The performance results for Case 2 are shown in Fig. \ref{fig-case2-multiplot-performance-0} for $d$=[3, 5, 7, 9] and Fig. \ref{fig-case2-multiplot-performance-1} for d=[11, 13, 15, 17]. 
            
            \begin{figure}[!ht]
                \centerline{\includegraphics[width=.75\columnwidth]{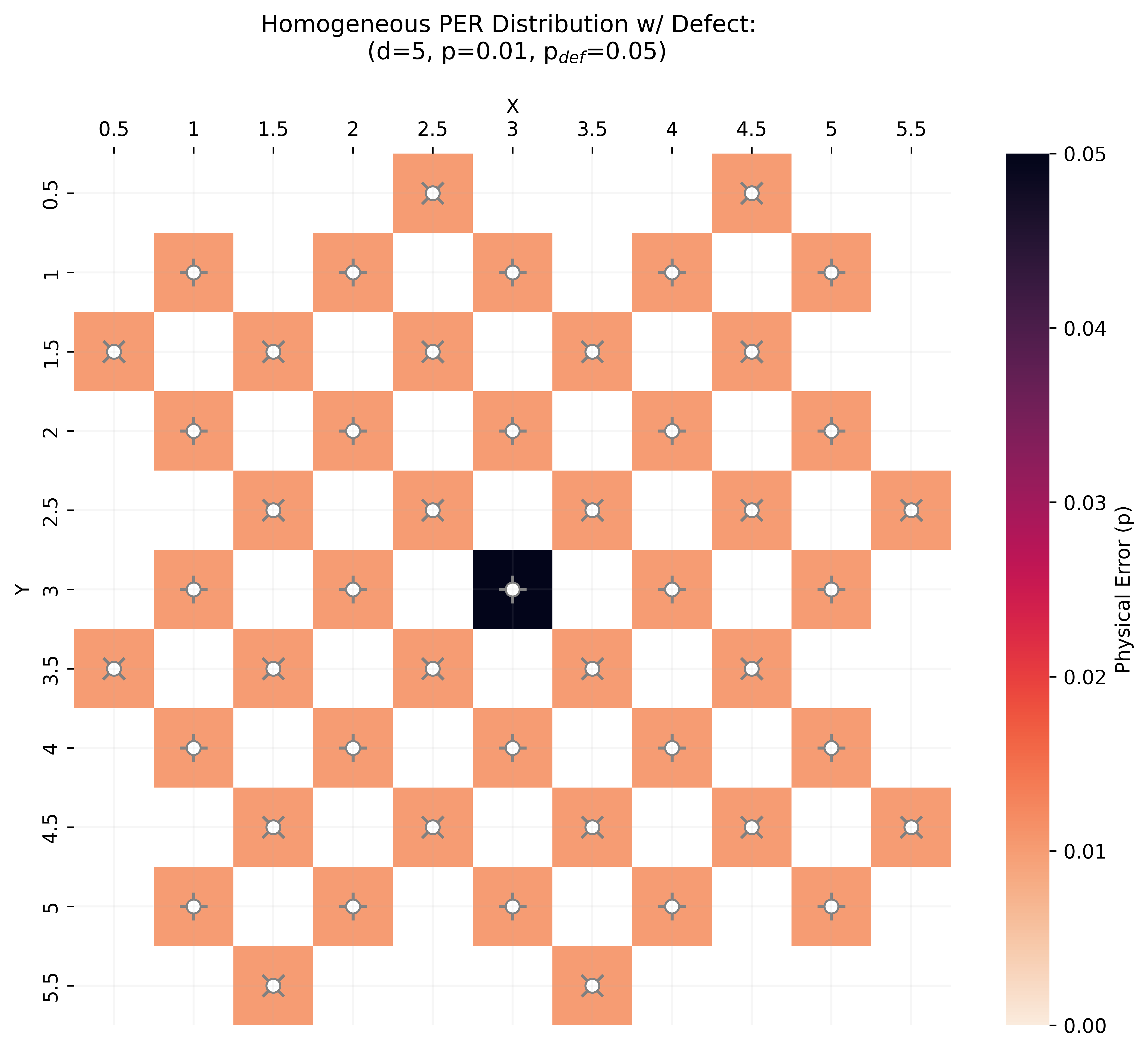}}
                \caption{ \textit{$($Case Study 2$)$} Heatmap for a homogeneous noise model on a d=5 code with a defect on the center data qubit. The \pdef shown in this heatmap is 0.05 to show this distribution, but significantly higher values \pdef are also used in the Case 2 experiments as well. In this heatmap, data qubits are denoted by \textbf{+} and measurement qubits are denoted with \textbf{X}.}
                \label{fig-case2-heatmap}
            \end{figure}
            \begin{figure}[!ht]
                \centerline{\includegraphics[width=1\columnwidth]{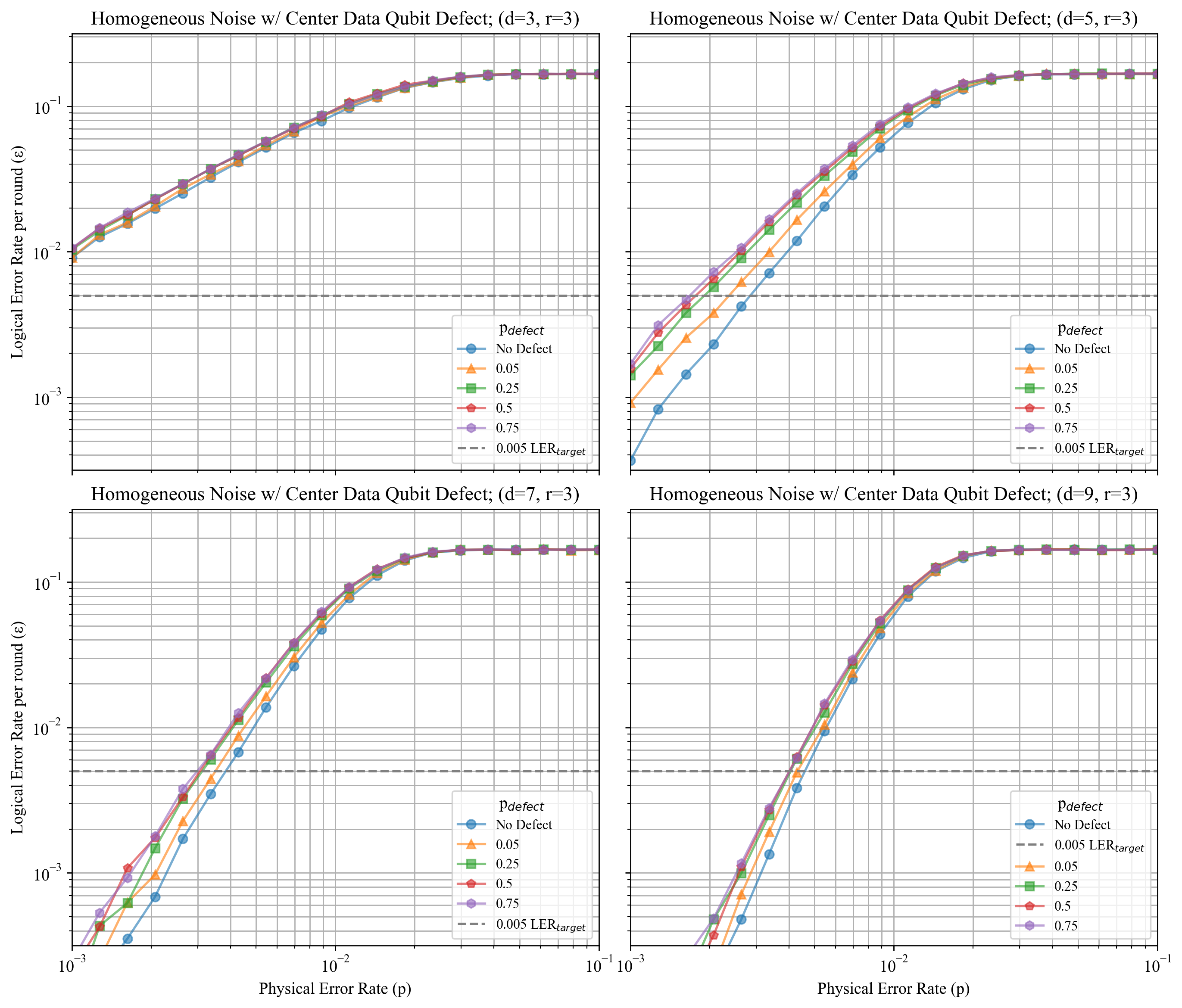}}
                \caption{ \textit{$($Case Study 2$)$} Performance of homogeneous noise model with outlier center data defect over d=3 through d=9. Each point in the plots represents a single simulated processor instance: for each fixed distance and $p_{defect}$, we generate one circuit per $p\in\{10^{-3+(\frac{2*k}{19}})|k\in\{0, 1, 2, ..., 19\}\}$ and estimate the logical error rate at that point from 50,000 shots of that circuit.}
                \label{fig-case2-multiplot-performance-0}
            \end{figure}
            \begin{figure}[!ht]
                \centerline{\includegraphics[width=1\columnwidth]{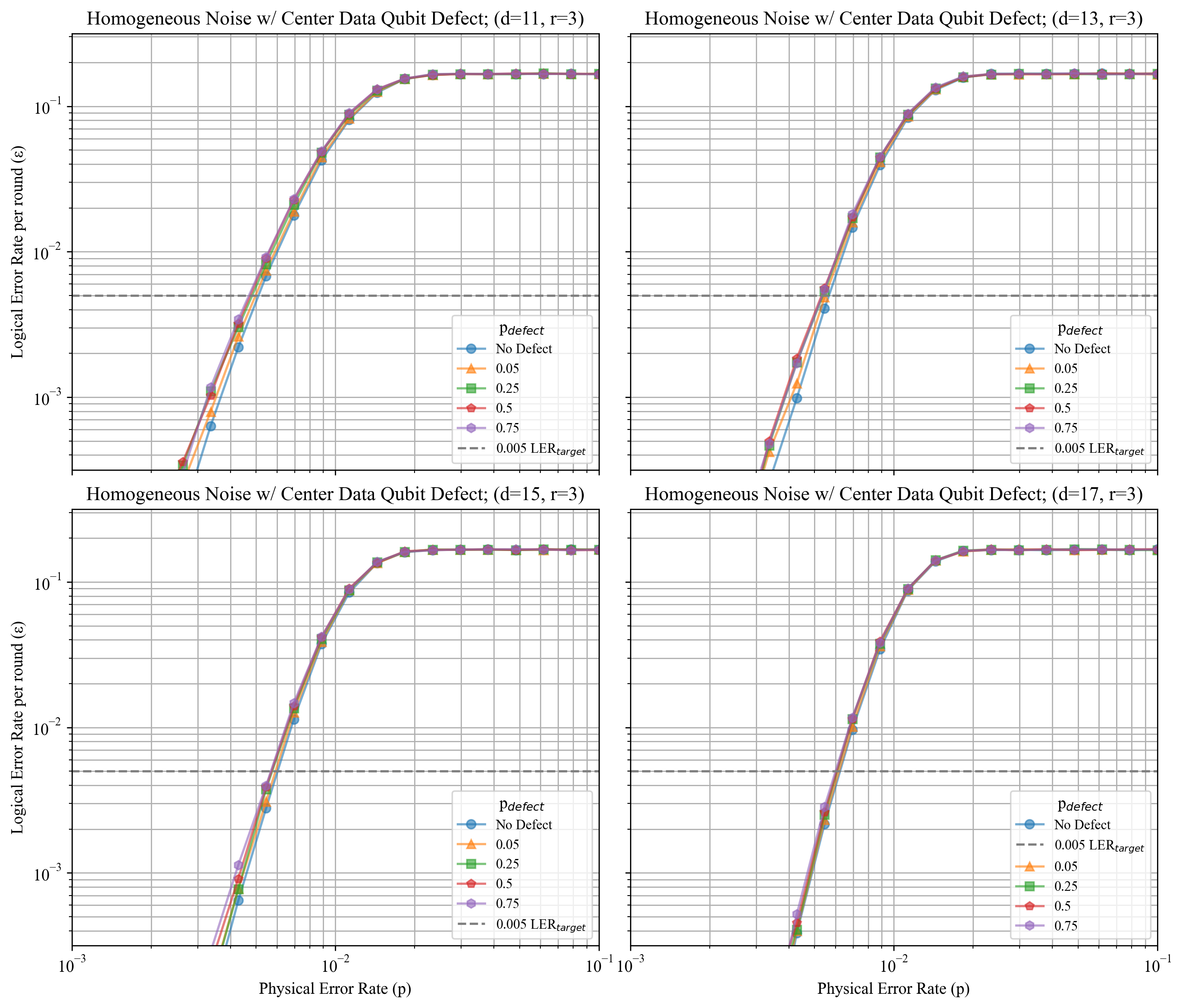}}
                \caption{ \textit{$($Case Study 2$)$} Performance of homogeneous noise model with outlier center data defect d=11 through d=17.}
                \label{fig-case2-multiplot-performance-1}
            \end{figure}
            
        In this experiment, all lattices across the sampled distances have a defect located at the data qubit in the center of the lattice with \pdef $\in\{0.05, 0.25, 0.5, 0.75\}$. At the upper end of the \pdef values sampled, the individual qubit is essentially completely unreliable, with roughly 75\% of operations on this qubit resulting in an erroneous computation. The results of this experiment (Fig. \ref{fig-case2-multiplot-performance-0}, Fig. \ref{fig-case2-multiplot-performance-1}, and Fig. \ref{fig-case2-xaxis-defect-value-over-code-distances}) show an incredibly strong resilience to the individual defect. Similar to that of the unilateral homogeneous model in Case 1 (Fig. \ref{fig-case1-performance}), all but the smallest code distance d=3 (Fig. \ref{fig-case2-multiplot-performance-0}) were able to meet the \ler{target}. In the worst case with \pdef = 0.75, the BAD for the d=17 code (Fig. \ref{fig-case2-multiplot-performance-1}) is $p\approx0.0058$. In comparing this to the d=17 performance in Case 1 (Fig. \ref{fig-case1-performance}), the maximally defective qubit only reduced our BAD by 3\%. For the minimum viable code distance in Case 2, d=5 (Fig. \ref{fig-case2-multiplot-performance-0}), the BAD is located at $p\approx0.0017$. 
        
        Specifically, for d=5 in Case 2 with the BAD for d=5 ($p\approx0.0029$) in Case 1 (Fig. \ref{fig-case1-performance}), the presence of the defect with \pdef = 0.75 resulted in a larger reduction to our acceptable boundary by about 50\%. As with physical noise tolerance, the results here suggest that under a homogeneous noise model, an increase in code distance also directly results in a significant increase in the resilience to individual defects with high levels of noise, which can be clearly seen in Fig. \ref{fig-case2-xaxis-defect-value-over-code-distances}.
        
            \begin{figure}[!ht]
            \centerline{\includegraphics[width=.8\columnwidth]{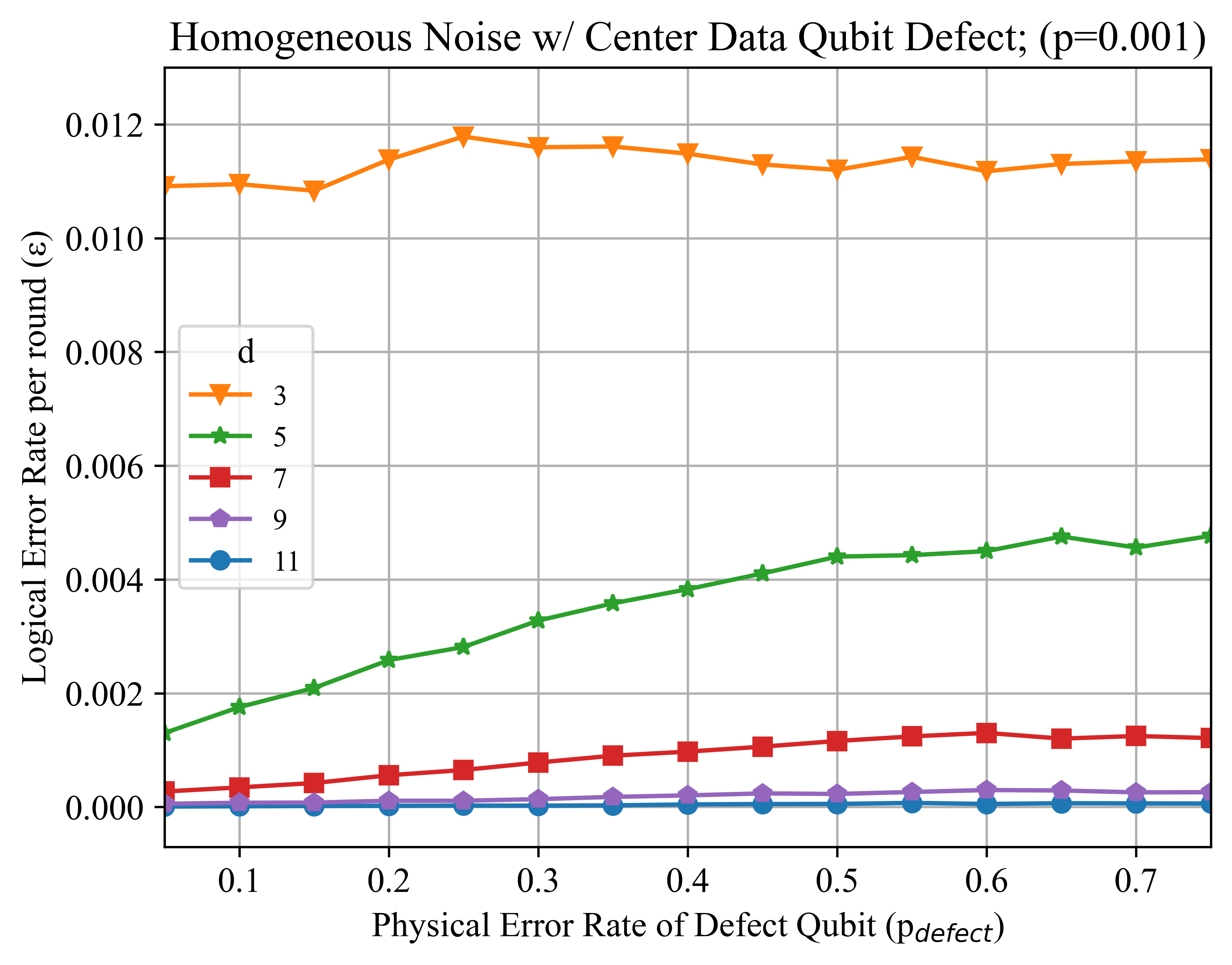}}
                \caption{ \textit{$($Case Study 2$)$} Logical error rate vs. defective physical qubit error rate for code distances d=[3, 5, 7, 9, 11]. To further examine how performance is affected across code distances, this figure shows a similar homogeneous noise model to that in Case 2, except that the noise for all other qubits is kept constant at \pvar{} = 0.001. As we increase our code distance, the lattice becomes increasingly resilient to the noise introduced by the defective data qubit, as expected. At distances of 9 or above, the performance is essentially unaffected by the presence of the defective qubit. Thus, this further suggests that the impact of highly defective qubits becomes negligible at a sufficiently large code distance. }
                \label{fig-case2-xaxis-defect-value-over-code-distances}
            \end{figure}
    }
    \subsection{Case 3: Heterogeneous (Gaussian) Noise}{
    \label{subsec-case-3}
        In the third case, we expand our noise model to a heterogeneous normal distribution of physical errors and examine the impact on the logical performance. A heatmap demonstrating a potential configuration for a d=5 heterogeneous noise model can be seen in Fig. \ref{fig-case3-heatmap}. The performance results for Case 3 are shown in Fig. \ref{fig-case3-multiplot-performance-0} for d=[3, 5, 7, 9] and Fig. \ref{fig-case3-multiplot-performance-1} for d=[11, 13, 15, 17]. 
            \begin{figure}[!ht]
                \centerline{\includegraphics[width=.75\columnwidth]{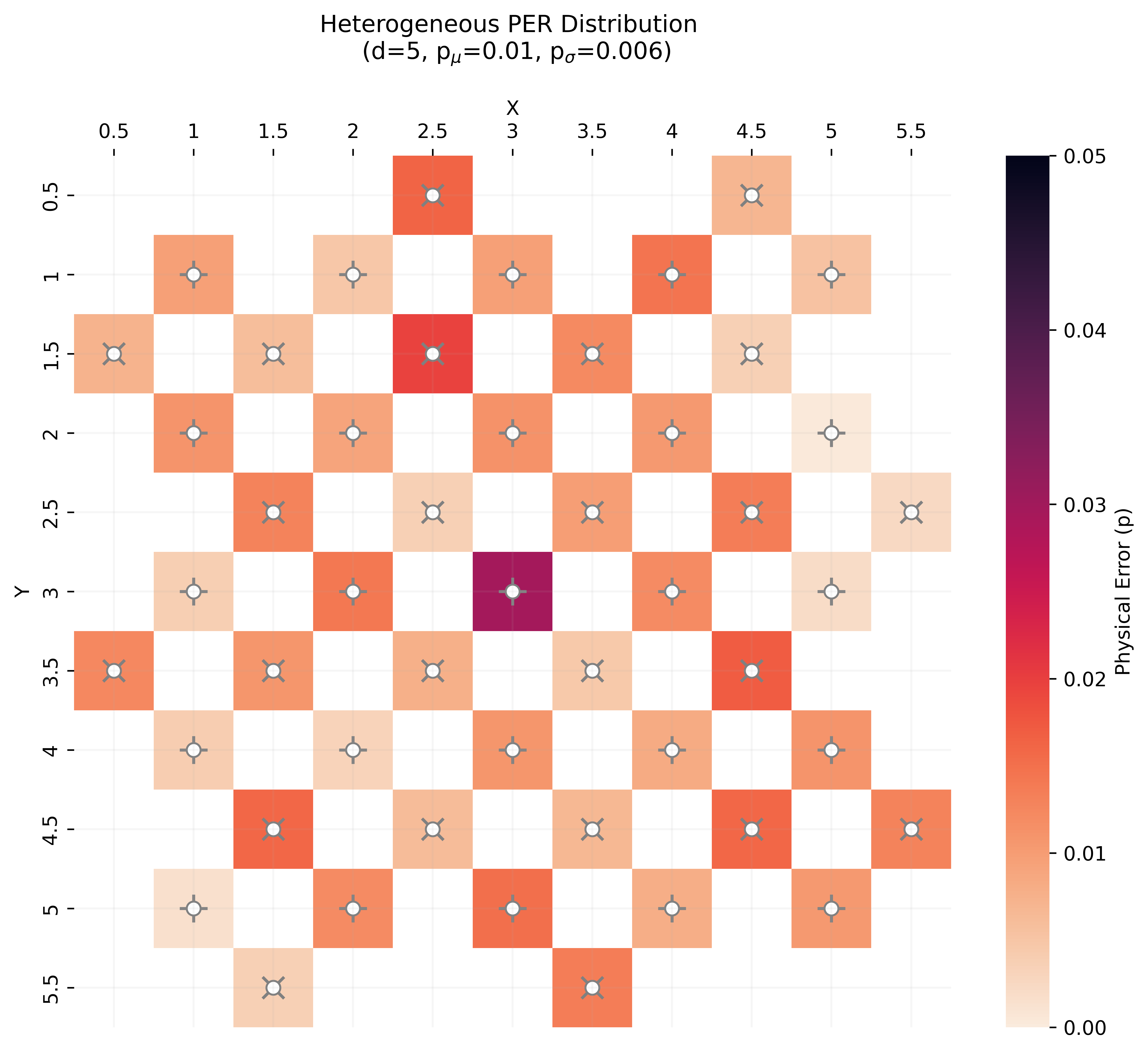}}
                \caption{ \textit{$($Case Study 3$)$} Heatmap for a heterogeneous noise model on a d=5 code. Data qubits are denoted by \textbf{+} and measurement qubits are denoted with \textbf{X}.}
                \label{fig-case3-heatmap}
            \end{figure}
            \begin{figure}[!ht]
                \centerline{\includegraphics[width=1\columnwidth]{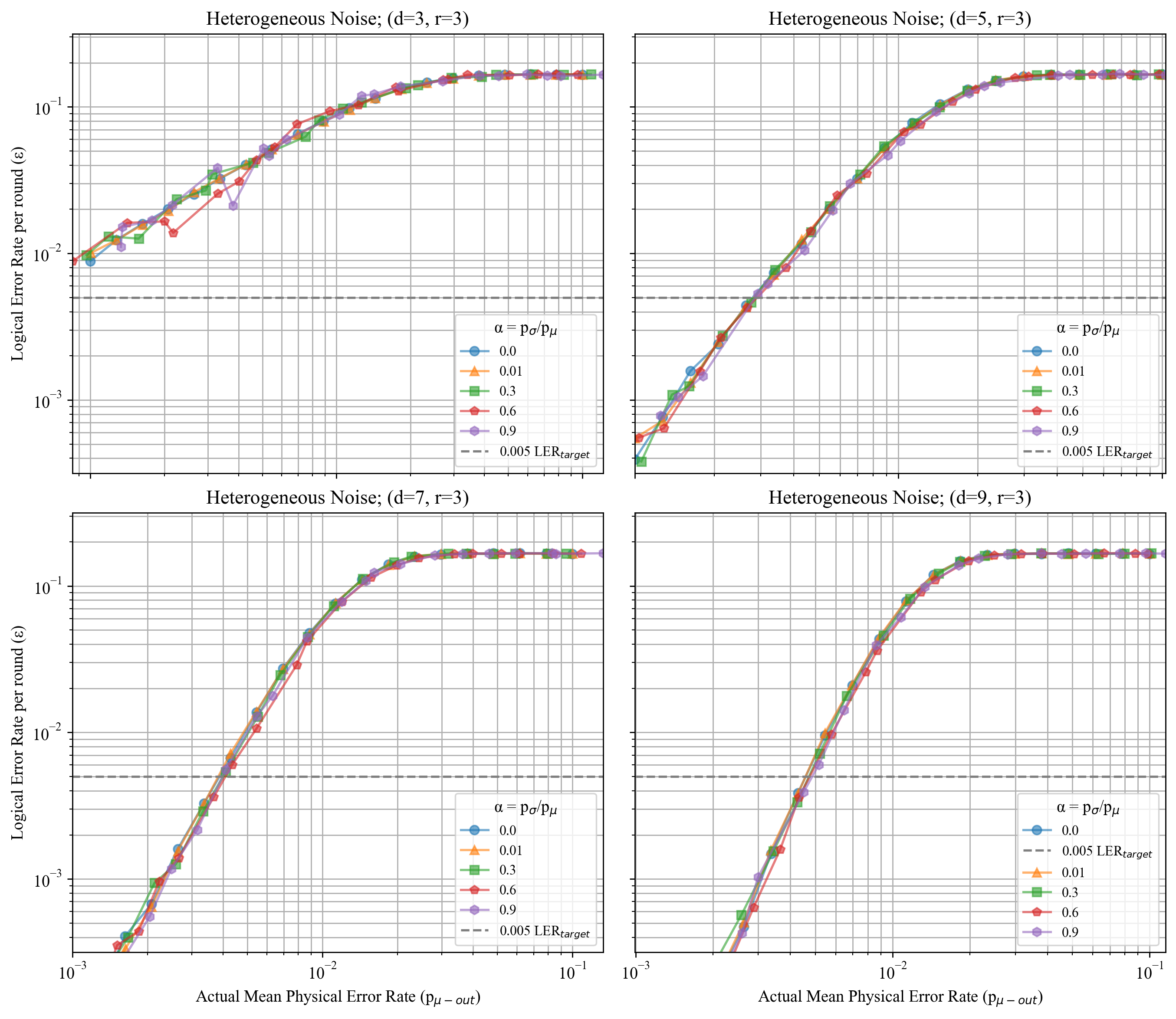}}
                \caption{ \textit{$($Case Study 3$)$} Performance of heterogeneous noise model for distances 3 through 9. The blue line ($\alpha$ = 0) represents a unilateral homogeneous noise model. Each point in the plots represents a single simulated processor instance: for each fixed distance and $\alpha$, we generate one circuit per $p_{\mu}\in\{10^{-3+(\frac{2*k}{19}})|k\in\{0, 1, 2, ..., 19\}\}$ and estimate the logical error rate at that point from 50,000 shots of that circuit.}
                \label{fig-case3-multiplot-performance-0}
            \end{figure}
            \begin{figure}[!ht]
                \centerline{\includegraphics[width=1\columnwidth]{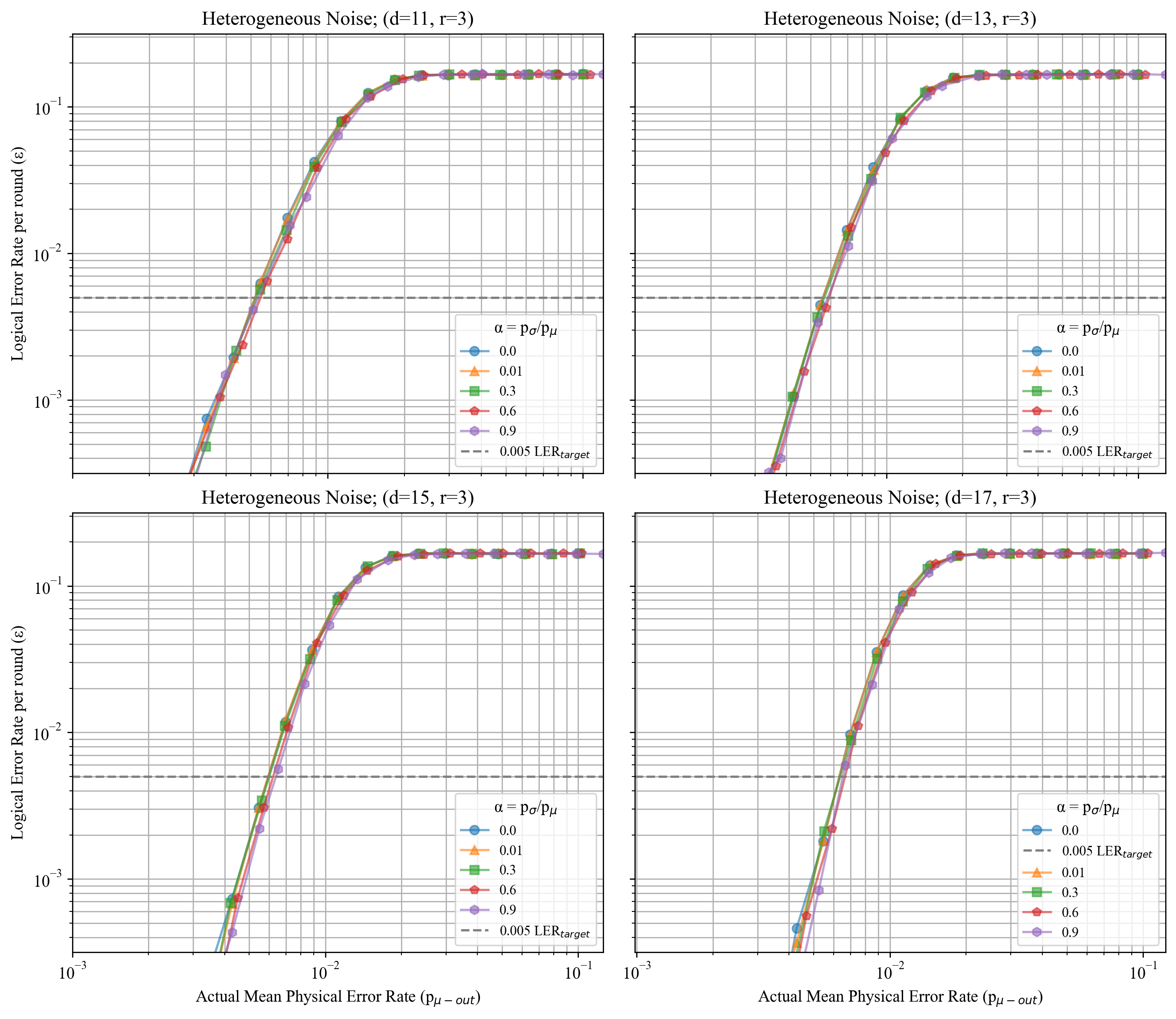}}
                \caption{ \textit{$($Case Study 3$)$} Performance of heterogeneous noise model for distances 11 through 17. The blue line ($\alpha = 0.0$) represents a unilateral homogeneous noise model covered in Case 1.}
                \label{fig-case3-multiplot-performance-1}
            \end{figure}
        For this experiment, we simulate heterogeneity by assigning each qubit a $p_i$ randomly sampled from a truncated normal distribution with an input mean parameter, \pmu, and an input standard deviation, \psigma. All operations involving this qubit in this instance will use the $p_i$ following the same operational noise assignment detailed in Section \ref{subsec-operational-noise}. To examine the impact of this Gaussian model of heterogeneous noise on logical performance, we simulated $\alpha\in\{0.01, 0.3,0.6,0.9\}$, corresponding to normal distributions with a deviation of 1\% to 90\% of the \pmu. For each $\alpha$ value examined, we simulated their performance with 20 different \pmu values evenly distributed between $10^{-3}$ and $10^{-1}$ on a logarithmic scale. To prevent negative probabilities from being sampled, a truncated distribution is used. When a distribution is truncated, the input mean used to create the distribution, \pmu, can differ from the \textit{actual} mean, \pmuout, of the values sampled from the distribution. Thus, to accurately reflect any shifts between input and actual mean, the \pmuout is used as the x-axis in our performance plots. For clarity, if  $p_{q_i}$ is the randomly sampled physical noise for qubit $q_i$ and $n = 2d^2-1$, then the average mean of the sampled distribution is as follows:
        
        \begin{equation}
        p_{\mu-out} = \frac{\overset{\text{n-1}}{\underset{\text{i=0}}{\sum}}p_{q_i}}{n}
        \end{equation}
        Similar to Case 1 (Fig. \ref{fig-case1-performance}) and 2 (Fig. \ref{fig-case2-multiplot-performance-0} and Fig. \ref{fig-case2-multiplot-performance-1}), the performance of the heterogeneous noise is benchmarked against a uniform homogeneous distribution, which is equivalent to when the $\alpha$ = \psigma = 0.0. The non-monotonic behavior of the plotted data in Fig. \ref{fig-case3-multiplot-performance-0} and Fig. \ref{fig-case3-multiplot-performance-1} is a result of this randomly sampled distribution, in which some circuits have a higher distribution of physical noise placed on more detrimental locations, such as higher noise on measurement qubits or edge qubits of which have a greater impact on LER \cite{fowler_surface_2012}. However, the noise and coherence properties of individual superconducting qubits can change frequently \cite{kimErrorMitigationStabilized2025}. Thus, purely monotonic performance curves when utilizing superconducting qubits is not a guarantee. 
        
        This is most notable with d=3 in Fig. \ref{fig-case3-multiplot-performance-0} in which the \ler{round} appears to fluctuate more drastically as $\alpha$ increases. In a d=3, the proportion of 2-weight stabilizers on the edge of a lattice to 4-weight stabilizers in the bulk of the lattice is higher. With a higher proportion of low-weight stabilizers in d=3, the increased deviation leads to greater fluctuations, as shown for $\alpha=0.9$ and $\alpha=0.6$ in Fig. \ref{fig-case3-multiplot-performance-0}.
    }
    \subsection{Case 4: Heterogeneous (Gaussian) Noise with Outlier Defect}{
    \label{subsec-case-4}
        In the fourth case, we examine performance changes under a heterogeneous noise model in the presence of a center data-defect qubit. A heatmap demonstrating a potential configuration for a d=5 heterogeneous noise model with a center defect can be seen in Fig. \ref{fig-case4-heatmap}. The performance results for Case 3 are shown in Fig. \ref{fig-case4-multiplot-performance-0} for d=[3, 5, 7, 9] and Fig. \ref{fig-case4-multiplot-performance-1} for d=[11, 13, 15, 17]. 
            \begin{figure}[!ht]
                \centerline{\includegraphics[width=.75\columnwidth]{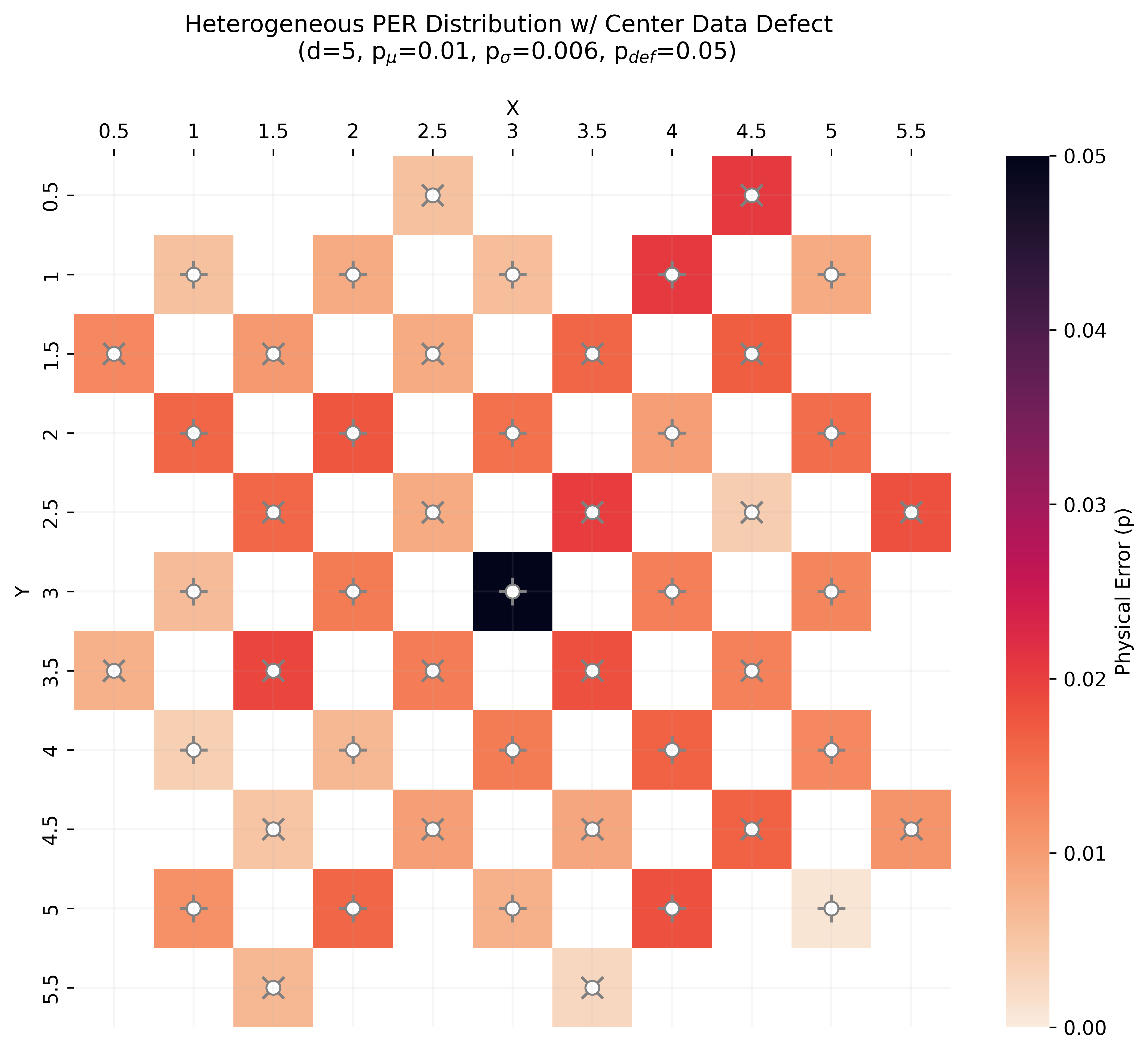}}
                \caption{ \textit{$($Case Study 4$)$} Heatmap for a heterogeneous noise model on a d=5 code with an outlier center data qubit. In this heatmap, data qubits are denoted by \textbf{+} and measurement qubits are denoted with \textbf{X}.}
                \label{fig-case4-heatmap}
            \end{figure}
            \begin{figure}[!ht]
                \centerline{\includegraphics[width=1\columnwidth]{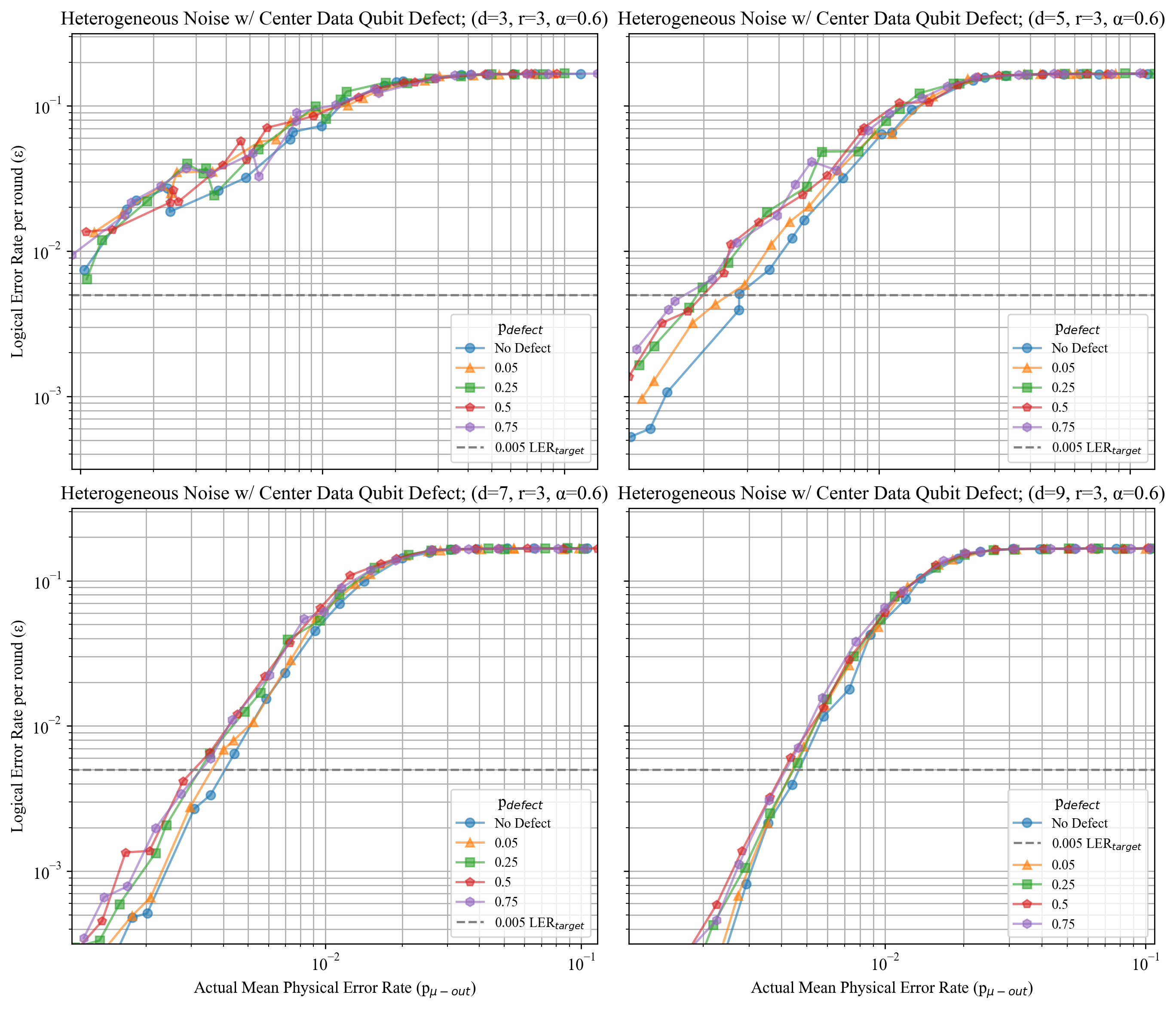}}
                \caption{ \textit{$($Case Study 4$)$} Performance of heterogeneous noise model with center data defect for distances 3 through 9. In Case 4, we simulated the performance of heterogeneous noise resembling a Gaussian distribution with a single defect at the center data qubit with a \pdef in the range of 0.05 to 0.75. In all simulations in Case 4, the $\alpha$ parameter, which determines the deviation as a percentage of the input mean, is held constant at $\alpha=0.6$. The blue line represents a baseline heterogeneous noise model with no defects present, which is the same as the $\alpha=0.6$ simulations in Case 3. Each point in the plots represents a single simulated processor instance: for each fixed distance and $p_{defect}$, we generate one circuit per $p_{\mu}\in\{10^{-3+(\frac{2*k}{19}})|k\in\{0, 1, 2, ..., 19\}\}$ and estimate the logical error rate at that point from 50,000 shots of that circuit.}
                \label{fig-case4-multiplot-performance-0}
            \end{figure}
            \begin{figure}[!ht]
                \centerline{\includegraphics[width=1\columnwidth]{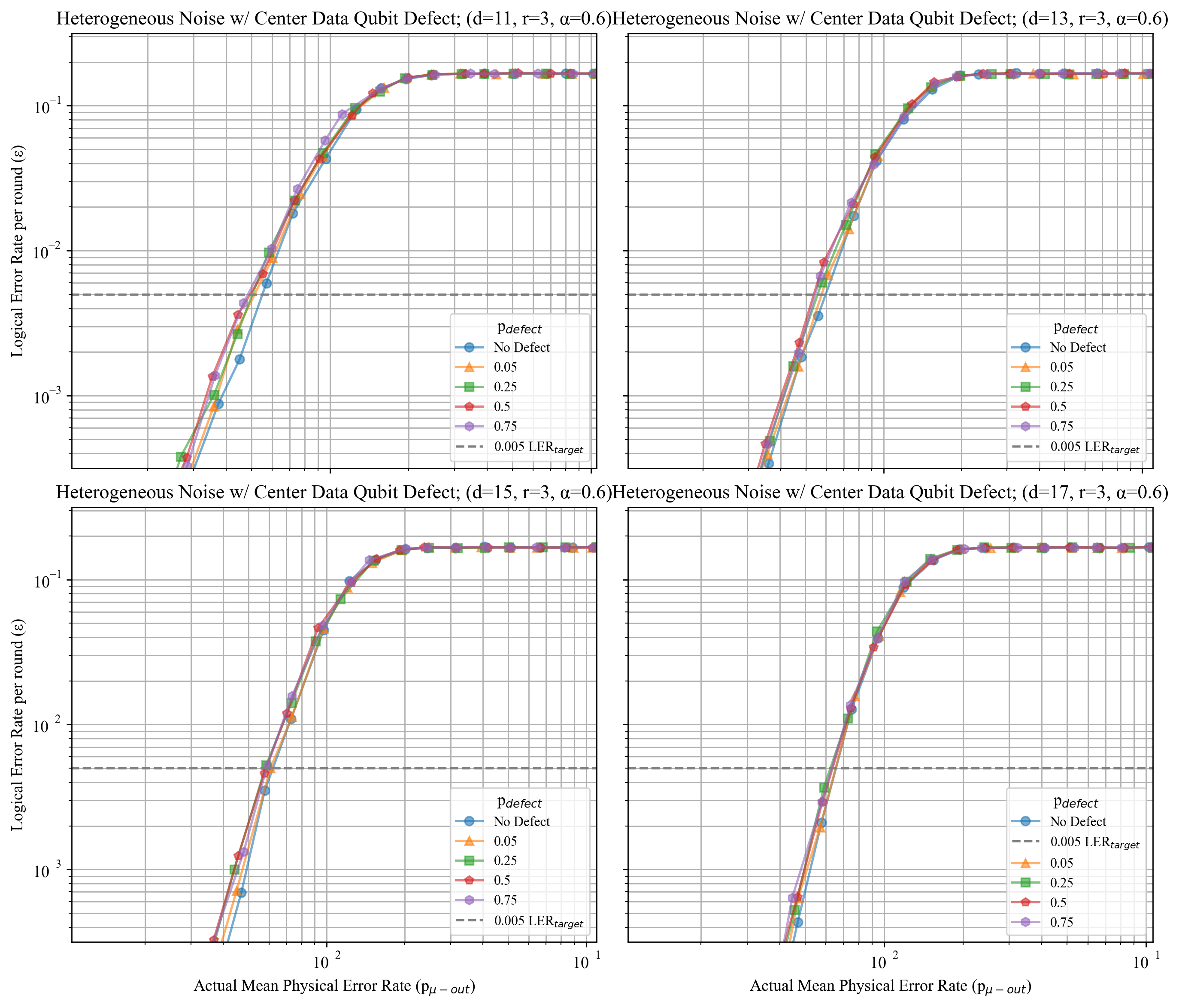}}
                \caption{ \textit{$($Case Study 4$)$} Performance of heterogeneous noise model with center data defect for distances 11 through 17. The blue line represents a baseline heterogeneous noise model with no defects present.}
                \label{fig-case4-multiplot-performance-1}
            \end{figure}
        The location of the defect (center data qubit) and \pdef $\in\{0.05, 0.25, 0.5, 0.75\}$ is identical to the setup in Case 2, with the exception that the noise of all other qubits is normally distributed rather than constant. As with the setup of the experiments in Case 3, we use the actual mean of the physical noise that was randomly sampled, \pmuout, as the horizontal axis. Additionally, like in Case 2, we want to assess how increasing the noise of all other qubits in the presence of a defect impacts the LER. Thus, our \pmuout does not include the \pdef value as it is not sampled from the distribution itself. For clarity, in Case 4, the \pmuout is as follows:
        
        \begin{equation}
        p_{\mu-out} = \frac{\overset{\text{n-1}}{\underset{\text{i=0}}{\sum}}p_{q_i}-p_{def}}{n-1}
        \end{equation}
        
        For this experiment, we selected a single $\alpha$ used in Case 3, $\alpha=.6$, and used the same \pdef values that were selected in Case 2. Interestingly, the performance trends shown in Case 4 (Fig. \ref{fig-case4-multiplot-performance-0} and Fig. \ref{fig-case4-multiplot-performance-1}) follow a pattern similar to that observed in Case 2 (Fig. \ref{fig-case2-multiplot-performance-0} and Fig. \ref{fig-case2-multiplot-performance-1}) in which higher distances show a strong resilience to increasing magnitudes of \pdef. 
    }
}
\section{Discussion} {
\label{sec-discussion}
    Our experiments aimed to highlight how heterogeneity affects performance on the SC, and the results indicate that the presumed model of noise greatly impacts the purported performance of the SC. We proposed in this work that, in order to accurately model SC on superconducting qubits, accepting some level of heterogeneity is necessary due to the inherent non-uniform noise exhibited by superconducting qubits. The aim of our work is thus to provide a preliminary accounting of how the performance of heterogeneous distributions directly affects the LER. As shown in our experiments, the presence of a single defective qubit with an error rate as high as 75\% significantly affects the LER only at small code distances and becomes increasingly negligible as the code distance increases. 
    
    With regard to the impact of individual defects, the performance data for both a single defect in the presence of heterogeneous and homogeneous noise models show that the defective qubits impact is only non-negligible at distances at or closely above the minimum viable code distance (the distance at which the target LER is reachable).
    
    In the performance experiments for Case 2 (Fig. \ref{fig-case2-multiplot-performance-0} and Fig. \ref{fig-case2-multiplot-performance-1}), the minimum viable code distance occurs at d=5. At d=3, the target LER is not reachable even when there are no defects present. Although the performance at d=3 is insufficient, the performance across the increasing \pdef noise is nearly identical, indicating that this defect has little impact on the LER. At d=5, the trend lines suddenly deviate, and the impact of higher \pdef noise on performance is noticeable. At d=5 through d=9, this performance deviation between \pdef noise begins to converge. At d=11 and above, the impact of an increasingly defective qubit is nearly negligible again. What we can extrapolate from this is that there exists a sort of ``Goldilocks zone" in which the detrimental impact of increasing noise on the defective qubit is evident. Within this "Goldilocks zone," choosing to drop out or mitigate the defective qubit with large \pdef values could result in noticeable performance gains. At d=11 and above, however, the impact of this defective qubit, even at p=0.75, is negligible and thus implementing a mitigation technique incurs additional cost, such as an increase in physical overhead \cite{yin_surf-deformer_2024, smith2022scaling} or circuit depth \cite{debroy_luci_2024}, but results in no performance gains. Ideally, when making design decisions, techniques that incur additional costs should only be used when there is a noticeable performance gain. In this way, our research suggests that treating defect mitigation should be viewed as an optimization problem in which we aim to only drop out qubits in which the resulting performance gain is non-negligible in light of the cost incurred by the mitigation method.
    
    Regarding the impact of heterogeneous noise that resembles a Gaussian distribution, our results show that the LER is not greatly affected by increasing deviation. In comparing the performance curves in Cases 1 with static \textit{p} and Cases 3 with normally distributed noise, we see nearly identical trends, indicating the increased deviation had a minimal impact on the LER. In Case 4, the same defect configuration (center data qubit) is tested under the heterogeneous model with a fixed $\alpha=0.6$. In the performance experiments conducted (Fig. \ref{fig-case4-multiplot-performance-0} and Fig. \ref{fig-case4-multiplot-performance-1}), the defect noise reflects a significant outlier from the remainder of the comparatively small normally distributed noise. These results closely resemble the trends discussed in Case 2. When comparing the insights of Case 1 vs. Case 3 and Case 2 vs. Case 4, our results indicate that while evenly distributed distributions may not impact LER, outlier defects much higher than the average physical noise do. The simulations used 3 rounds across all distances, which was done to enable rapid experimenting over many configurations and distances while we were exploring these preliminary models of noise. We performed the same experiments for each case study with rounds=distance. While higher \pdef values became slightly more noticeable impact on LER at higher distances, the trends discussed for all cases involving both defects and impact of Gaussian noise were the same.
    
    It appears that if noise is normally distributed, the slightly higher fidelity qubits essentially cancel out their slightly noisier counterparts. However, this balancing requires that for every bad qubit in the lattice, there is an equally good qubit, which is an unrealistic assumption. Further, the presence of a significant defect cannot be balanced and thus has a noticeable impact on the LER. After all, it is much easier to fabricate a noisy qubit than it is to fabricate a perfect qubit. Ultimately, a heterogeneous noise model based on a Gaussian distribution does not appear to capture the complexity or impact of real heterogeneous noise, nor does it accurately reflect what we observe on real devices, further arguing that real quantum noise on superconducting devices cannot be fully represented as Gaussian. 
}
\section{Future Work} {
\label{sec-future-work}
    Overall, the results of our case studies concur with prior work examining the impact of individual defects and normally distributed heterogeneous noise \cite{carrollSubsystemSurfaceCompass2024}, and expand these insights across a wide range of distances, outlier defects, and distribution deviations. Going forward, we believe that examining the impacts of heterogeneity on SC is vital as both physical devices and QEC techniques advance toward fault-tolerant quantum computing. Building upon the tool and results presented here, we believe there are many areas in which this work could be expanded. A particular area of further development would be to expand the simplified noise model that we use to create our circuits, such as introducing additional scaling factors to more accurately represent the different operational error rates \cite{mohseni_how_2025}. Second, the experiments focused primarily on testing individual defects on a single data qubit, but expanding these experiments to encompass how additional configurations of defect characteristics affect performance is an area for future work. Identifying how these defects accumulate would provide further insight into when deformation techniques are absolutely necessary. The experimental results presented in our research demonstrate preliminary evidence showing the existence of BADs as well as the disparity between viewing the SC under a homogeneous model vs. a heterogeneous model. If a primary goal of simulation is to optimize performance for computation on real devices, restricting examinations of the SC to only homogeneous noise models can lead to vastly different design outcomes, such as picking optimal code distances to meet a target logical error rate, that fail to fully utilize the already limited physical resources available in present and near-term quantum hardware.

    Secondly, our current analysis in the distributed heterogeneous case studies (Sec. \ref{subsec-case-3} and \ref{subsec-case-4}) focuses on single-processor samples for each parameter setting (e.g., \pdef or $\alpha$), where each plotted point corresponds to one simulated processor instance with many shots, rather than an ensemble average over independently resampled processors. As a result, we do not report statistical error bars across heterogeneous instances in this work, due in part to computational resource constraints for this initial study. Another natural direction for future work is to extend these experiments to an ensemble regime in which, for each set of parameters, many independent processors are generated and averaged. Such ensemble-based studies would enable us to quantify variability more adeptly across different heterogeneous noise landscapes. 

    Finally, these results suggest that modeling heterogeneity with a Gaussian distribution is a poor proxy for the kinds of imbalance observed on superconducting devices. When noise on a lattice is approximately normally distributed, the slightly better-than-average qubits effectively counterbalance the slightly worse ones within a logical patch, and the resulting LER closely resembles that of an idealized homogeneous lattice with a physical error rate equal to the mean. In this sense, a Gaussian model collapses back to the homogeneous assumption, with the \pmu playing the same role as a uniform $p$ in the standard surface-code scaling relation in \ref{sec-homogeneous-vs-heterogeneous-assumptions}. However, this same observation also points to a useful design goal: when selecting logical patches and code distance on a fixed, heterogeneous device, one might aim to partition the physical qubits so that each logical patch's noise distribution is as close to balanced (i.e., approximately Gaussian) as possible, thereby allowing high-fidelity qubits to offset noisier ones and achieve near-homogeneous performance. Since empirical studies indicate that real-device noise is neither perfectly Gaussian nor perfectly balanced \cite{sungNonGaussianNoiseSpectroscopy2019}, this further motivates future work on noise models and mapping strategies that explicitly capture skewed, multi-modal, or otherwise imbalanced error landscapes rather than relying on symmetric Gaussian assumptions.
}
\section{Conclusion} {
    In this work, we explore the impact of heterogeneous noise on the SC's performance. This work is among the first to study the physical error rate of device qubits as a distribution and the first to provide a framework and preliminary toolset for rapidly evaluating how configurations of non-identical noise affect the LER in rotated SCs. Using the BADs framework, we explore the impact of both outlier qubits and normally distributed physical errors on SC LER. As a result of our work, we discovered that one bad qubit does not necessarily spoil the bunch. Furthermore, we demonstrate that if heterogeneous noise follows a Gaussian distribution, the SC LER scales only with the mean of the SC lattice, not its deviation. This ultimately indicates that modeling heterogeneous noise as a Gaussian distribution is functionally equivalent to treating it as homogeneous, further emphasizing the need to develop more complex models that capture the effects of superconducting qubit deviation. To conclude, this work motivates the inclusion of BADs in QEC analysis based on the SC, and we hope that the software framework developed here will open the door to design space exploration, enabling shorter timelines to fault-tolerant quantum computing.
}
\section{Declarations}{
    \subsection{Acknowledgments}
    The authors are grateful to Malcolm Carroll and Andrew Cross for productive discussions on heterogeneous noise modeling and for their thoughtful feedback, which significantly strengthened this work.
    
    \subsection{Data Availability}
    The data and figures generated for this paper were created from our code and can be found on our GitHub repository at: \url{https://github.com/JacobSPalmer/quantum-ddq-toolkit/tree/icrc-25}. Additionally, all data used to generate the specific figures shown in the manuscript can be found here: \url{https://github.com/JacobSPalmer/quantum-ddq-toolkit/tree/icrc-25/supplementary-materials}.
    
    \subsection{Author Contributions}
    J. S. P. led code development and experiment implementations. K.N.S. advised project direction, and both authors contributed to the analysis and writing of the manuscript.
    
    \subsection{Funding}
    No funding was received in relation to this study. (Funding: Not Applicable) 
    
    \subsection{Competing Interests}
    The authors of this paper declare no competing interests.
}
\printbibliography

\end{document}